\documentclass[ aip, jmp, reprint,]{revtex4}
\usepackage{amsfonts}
\usepackage{url}
\usepackage{amssymb}
\usepackage{amsmath}
\usepackage{graphicx}
\usepackage{dcolumn}
\usepackage{bm}

\begin{document}

\title[]{\textbf{Hamiltonian structure of classical }$N$-\textbf{body
systems \\ of finite-size particles subject to EM interactions}}
\author{Claudio Cremaschini}
\affiliation{International School for Advanced Studies (SISSA) and INFN, Trieste, Italy}
\affiliation{Consortium for Magnetofluid Dynamics, University of Trieste, Italy}
\author{Massimo Tessarotto}
\affiliation{Department of Mathematics and Informatics, University of Trieste, Italy}
\affiliation{Consortium for Magnetofluid Dynamics, University of Trieste, Italy}
\date{\today }

\begin{abstract}
An open issue in classical relativistic mechanics is the consistent
treatment of the dynamics of classical $N$-body systems of
mutually-interacting particles. This refers, in particular, to charged
particles subject to EM interactions, including both binary and self
interactions (\emph{EM-interacting }$N$-\emph{body systems}). The correct
solution to the question represents an overriding prerequisite for the
consistency between classical and quantum mechanics. In this paper it is
shown that such a description can be consistently obtained in the context of
classical electrodynamics, for the case of a $N$-body system of classical
finite-size charged particles. A variational formulation of the problem is
presented, based on the $N$-body hybrid synchronous Hamilton variational
principle. Covariant Lagrangian and Hamiltonian equations of motion for the
dynamics of the interacting $N$-body system are derived, which are proved to
be delay-type ODEs. Then, a representation in both standard Lagrangian and
Hamiltonian forms is proved to hold, the latter expressed by means of
classical Poisson Brackets. The theory developed retains both the covariance
with respect to the Lorentz group and the exact Hamiltonian structure of the
problem, which is shown to be intrinsically non-local. Different
applications of the theory are investigated. The first one concerns the
development of a suitable Hamiltonian approximation of the exact equations
that retains finite delay-time effects characteristic of the binary and self
EM interactions. Second, basic consequences concerning the validity of Dirac
generator formalism are pointed out, with particular reference to the
instant-form representation of Poincar\`{e} generators. Finally, a
discussion is presented both on the validity and possible extension of the
Dirac generator formalism as well as the failure of the so-called Currie
\textquotedblleft no-interaction\textquotedblright\ theorem for the
non-local Hamiltonian system considered here.
\end{abstract}

\maketitle



\section{Introduction}

In classical physics the formulation of the Hamiltonian mechanics of $N$%
-body systems composed of interacting particles is still incomplete. This
includes, in particular, the case of charged particles acted on by an
externally-prescribed EM field as well as binary and self EM interactions.
Indeed, based on general relativity (or special relativity, as appropriate
in the case of a flat Minkowski space-time) as well as quantum mechanics,
common prerequisites for a dynamical theory for such systems should be:

\emph{Prerequisite \#1}: its covariance with respect to arbitrary local
coordinate transformations. In the context of special relativity this
requirement reduces to the condition of covariance with respect to the
Lorentz group.

\emph{Prerequisite \#2}: the inclusion of both retarded and local
interactions.

\emph{Prerequisite \#3}: the consistency with the Einstein causality
principle.

\emph{Prerequisite \#4}: the validity of the Hamilton variational principle,
yielding a set of equations of motion for all the $N$ particles of the $N$%
-body system.

\emph{Prerequisite \#5}: the existence of a \emph{Hamiltonian structure}.

As clarified below, all of these statements should be regarded as intrinsic
properties of classical $N$-body systems which are characterized by
non-local, i.e., retarded causal interactions, like those associated with EM
fields \cite{Jackson}. In particular, requirements \#4 and \#5 involve the
assumptions that the equations of motion of a generic $N$-body system of
this type should admit both Lagrangian and Hamiltonian variational
formulations, obtained by means of a Hamilton variational principle, as well
as a \emph{standard Hamiltonian form, }i.e., a set $\left\{ \mathbf{x}%
,H_{N}\right\} $ with the following properties:

a) $\mathbf{x}=\left( \mathbf{x}^{\left( i\right) },i=1,N\right) $ is a
super-abundant canonical state, with $\mathbf{x}^{\left( i\right) }$
denoting an appropriate $i$-th particle canonical state;

b) $H_{N}$\ (to be referred to as \emph{system Hamiltonian}) is a suitably
regular function. In view of prerequisites \#2 and \#3, we expect $H_{N}$ to
be prescribed in terms of a \emph{non-local phase-function} of the form $%
H_{N}(\mathbf{x},\left[ \mathbf{x}\right] )$, $\mathbf{x}$ and\textbf{\ }$%
\left[ \mathbf{x}\right] $ denoting respectively local and non-local
dependences;

c) for all particles $i=1,N$ belonging to the $N$-body system, the
variational equations of motion must admit the \emph{standard Hamiltonian
form }expressed\emph{\ }in terms of the Poisson brackets with respect to the
system Hamiltonian, namely:%
\begin{equation}
\frac{d\mathbf{x}^{\left( i\right) }}{ds_{\left( i\right) }}=\left[ \mathbf{x%
}^{\left( i\right) },H_{N}\right] .  \label{STANDARD HAMILTONIAN FORM}
\end{equation}%
Here the notation is standard. Thus, $\left( s_{\left( 1\right)
},...s_{\left( N\right) }\right) $ and $\left[ \eta ,\xi \right] $ $\equiv %
\left[ \eta ,\xi \right] _{\left( \mathbf{x}\right) }$ are respectively the
particles proper times and the \textit{local Poisson brackets} (PBs). The
latter are defined in terms of the super-abundant canonical state $\mathbf{x}
$ as
\begin{equation}
\left[ \eta ,\xi \right] =\left( \frac{\partial \eta }{\partial \mathbf{x}}%
\right) ^{T}\cdot \underline{\underline{\mathbf{J}}}\cdot \left( \frac{%
\partial \xi }{\partial \mathbf{x}}\right) ,  \label{PB}
\end{equation}%
with all components of $\mathbf{x}$ to be considered independent (i.e., $%
\mathbf{x}$ as \emph{unconstrained}). Furthermore, $\underline{\underline{%
\mathbf{J}}}$ is the canonical Poisson matrix \cite{Goldstein}, while $\eta (%
\mathbf{x})$ and $\xi (\mathbf{x})$ denote two arbitrary smooth
phase-functions.

Evidently, the above prerequisites should be regarded as overriding
conditions for the transition from classical to quantum theory of the $N$%
-body dynamics to be possible. However, despite notable efforts (see for
example Dirac, 1949 \cite{Dirac}) the solution to the problem of fitting
them together has remained still incomplete to date, at least in the case of
systems of charged particles subject to EM interactions.

From the point of view of classical physics the reason is related to the
nature of EM interactions occurring in $N$-body systems. These can be
carried respectively both by external sources (unary interactions, due to
prescribed external EM fields) and by the particles themselves of the system
(internal interactions). The latter include both binary EM interactions
acting between any two arbitrary charged particles and the self EM
interaction, usually known as the EM radiation-reaction (RR; Dirac \cite%
{Dirac1938}, Pauli \cite{Pauli1958}, Feynman \cite{Feynman1988}). As earlier
pointed out (see Refs.\cite{Cr2011A} and \cite{Cr2011B}, hereon referred to
as Paper I and Paper II respectively) a rigorous treatment of the EM
self-interaction consistent with prerequisites \#1-\#5 can only be achieved
for extended classical particles, i.e., particles characterized by mass and
charge distributions with finite support. In particular, a convenient
mathematical model is obtained by assuming that these classical particles
are non-rotating and their mass and charge distributions are quasi-rigid in
their rest frames (see Paper I for a detailed discussion on this point). In
Papers I and II the dynamics of single extended particles (1-body problem)
in the presence of their EM self-fields was systematically investigated in
the context of classical electrodynamics, by means of a variational approach
based on a Hamilton variational principle. As a result, the Hamiltonian
description for isolated particles subject to the combined action of the
external and the self EM interaction has been established.

However, fundamental issues still remain unanswered regarding the analogous
formulation of a consistent dynamical theory holding for classical $N$-body
systems of finite-size charges subject to only EM interactions (\emph{%
EM-interacting }$N$-\emph{body systems}). In fact, it is well known that
traditional formulations of the relativistic dynamics of classical charged
particles are unsatisfactory, at least because of the following main
reasons.\

The first one is related to the approximations usually adopted in classical
electrodynamics for the treatments of RR phenomena. In most of previous
literature charged particles are regarded as point-like and the so-called
short delay-time ordering%
\begin{equation}
\epsilon \equiv \frac{t-t^{\prime }}{t}\ll 1
\label{SHORT DELAY-TIME ORDERING}
\end{equation}%
is assumed to hold, with $t$ and $t^{\prime }$ denoting respectively the
\textquotedblleft present\textquotedblright\ and \textquotedblleft
retarded\textquotedblright\ coordinate times, both defined with respect to a
suitable Laboratory frame. This motivates the introduction of asymptotic
approximations, both for the EM self\ 4-potential and the corresponding
self-force, which are based on power-series expansions in terms of the
dimensionless parameter $\epsilon $ and are performed in a neighborhood of
the present coordinate time $t$ (see related discussion in Paper II).
Nevertheless, previous approaches of this type have lead in the past\emph{\
to intrinsically non-variational and therefore non-Hamiltonian equations of
motion} \cite{Cr2011A,Cr2011B,Dorigo2008a}. These are exemplified by the
well-known LAD and LL RR equations,\ due respectively to Lorentz, Abraham
and Dirac (Lorentz, 1985 \cite{Lorentz}; Abraham, 1905 \cite{Abraham1905}
and Dirac \cite{Dirac1938}) and Landau and Lifschitz \cite{LL}. Both
features make these treatments incompatible with the physical prerequisites
indicated above.

The second motivation arises in reference to the so-called \textquotedblleft
no-interaction\textquotedblright\ theorem proposed by Currie \cite%
{Currie1963,Currie1963B}. According to its claim, isolated $N$-body systems,
formed by at least two point particles, which are subject to mutual EM
binary interactions and in which the canonical coordinates are identified
with the space parts of the particle position 4-vectors, \emph{cannot define
a Hamiltonian system with manifestly covariant canonical equations of motion}%
. The correctness of such a statement has been long questioned (see for
example Fronsdal, 1971 \cite{Fronsdal} and Komar, 1978-1979 \cite%
{Komar1978,Komar1978a,Komar1978b,Komar1979}).\ In particular, the
interesting question has been posed whether \emph{the \textquotedblleft
no-interaction\textquotedblright\ theorem can actually be eluded in
physically realizable classical systems}. Interestingly, Currie approach is
based on the well-known generator formalism formulated originally by Dirac
(DGF; Dirac, 1949 \cite{Dirac}). Therefore, related preliminary questions
concern the conditions of validity of DGF itself and, in particular, whether
such an approach actually applies \emph{at all} to \emph{EM-interacting }$N$%
\emph{-body systems}.

\bigskip

\begin{center}
\textbf{Goals of the paper}
\end{center}

Put all the previous motivations in perspective, in this paper a systematic
solution to these issues is presented, for the case of EM-interacting $N$%
-body systems of classical finite-size charged particles. The theory is
developed in the framework of classical electrodynamics and special
relativity (i.e., assuming a flat Minkowski space-time) and is shown to
satisfy all the prerequisites indicated above (\#1-\#5).

Starting point is the determination of both Lagrangian and Hamiltonian
equations of motion for classical charged particles belonging to an
EM-interacting $N$-body system. By construction (see Section 2) the latter
can be considered as a system of smooth hard sphere, namely in which hard
collisions occurring between the particles, when their boundaries $\partial
\Omega _{\left( i\right) }$ (see below) come into contact, conserve each
particle angular momentum. In particular, for simplicity, in the following
all extended particles will be considered as acted upon only by EM
interactions, thus ignoring the effect of hard collisions on the $N$-body
dynamics. As in previous Papers I and II, where the 1-body problem was
investigated, the derivation is based on the variational formulation of the
problem. This requires, in particular, the determination of the appropriate
variational functionals required for the description of both binary and self
EM interactions. The resulting action functional is found to be expressed as
a line integral in terms of a suitable non-local variational Lagrangian, the
non-locality being associated both to the finite-extension of the charge
distributions and to delay-time effects arising in binary EM interactions.
Based on the Hamilton variational principle expressed in superabundant
variables, the resulting variational equations of motion (for the $N$-body
system) are then proved to be necessarily delay-type ODEs. As a consequence,
based on the definition of suitable effective Lagrangian and Hamiltonian
functions, a manifestly-covariant representation of these equations in both
standard Lagrangian and Hamiltonian forms is reached. The main goal of the
paper is then to show that these features allow a Hamiltonian structure $%
\left\{ \mathbf{x},H_{N}\right\} $ to be properly defined in terms of a
suitable superabundant canonical state $\mathbf{x}$ and a non-local system
Hamiltonian function $H_{N}.$ The result follows by noting that the
Hamiltonian equations in standard form admit also a representation in terms
of local PBs, defined with respect to the super-abundant canonical state $%
\mathbf{x}$.

A remarkable development concerns the construction of an approximation of
the Hamilton equations in standard form. This holds in validity of both the
short delay-time and large-distance orderings, namely under the same
asymptotic conditions usually invoked in the literature for the asymptotic
treatment of the RR problem.\ Based on the analogous approach developed in
Paper II, it is shown that a suitable $N$-\emph{body Hamiltonian
approximation} [of the exact problem] can actually be reached, which
preserves its Hamiltonian structure. In particular it is proved that, unlike
in the LAD and LL equations, the asymptotic approximation obtained in this
way keeps the variational character of the exact theory, retaining the
standard Lagrangian and Hamiltonian forms of the $N$-body dynamical
equations as well as the delay-time contributions arising from the various
EM interactions.

Further interesting conclusions are drawn concerning the validity of DGF in
the present context. This refers, in particular, to the so-called
instant-form representation of Poincar\`{e} generators for infinitesimal
transformations of the inhomogeneous Lorentz group. It is pointed out that
DGF in its original formulation only applies to \emph{local Hamiltonian
systems} and therefore is inapplicable to (the treatment of) the
EM-interacting $N$-body systems considered here. For definiteness, the
correct set of Poincar\`{e} generators, corresponding to the exact non-local
Hamiltonian structure $\left\{ \mathbf{x},H_{N}\right\} $ determined here,
together with their instant-form representation, are also provided. This
permits to develop a modified formulation of DGF, denoted as \emph{non-local
generator formalism}, which overcomes the previous limitations and is
applicable also to the treatment of non-local Hamiltonians in terms of \emph{%
essential} (i.e., \emph{constrained}) canonical variables.

Finally, on the same ground, the Currie \textquotedblleft
no-interaction\textquotedblright\ theorem is proved to be violated in any
case by the Hamiltonian structure, i.e., both by its \emph{exact realization
}$\left\{ \mathbf{x},H_{N}\right\} $ and\emph{\ }its\emph{\ asymptotic
approximation}. Counter-examples which overcome the limitations stated by
the \textquotedblleft no-interaction\textquotedblright\ theorem are
explicitly provided. In particular, the purpose of this discussion is to
prove that indeed a standard Hamiltonian formulation for the $N$-body system
of EM-mutually-interacting charged particles can be consistently obtained.
The main cause of the failure of the Currie theorem is identified in the
conditions of validity of DGF on which the proof of the theorem itself is
based.

\bigskip

\bigskip

\begin{center}
\textbf{Scheme of the presentation}
\end{center}

The paper is organized as follows. In Section 2 the expressions for the EM
current density and self 4-potential for extended charged particles are
recalled. In Section 3 the $N$-body action integral is constructed, based on
the results of Papers I and II and focusing in particular on the derivation
of the action integral for the EM binary interactions. Section 4 deals with
the dynamical equations of motion in Lagrangian form for the interacting
particles, derived from the $N$-body hybrid synchronous Hamilton variational
principle (THM.1). A standard Lagrangian form of the same equations is
obtained, by introducing an effective non-local Lagrangian function
(Corollary to THM.1). In Section 5 the corresponding non-local Hamiltonian
theory is developed in terms of a variational Hamiltonian function (THM.2).
A standard Hamiltonian form of the $N$-body dynamical equations is
constructed and the associated non-local Hamiltonian structure is proved to
exist (Corollary to THM.2). General implications of the non-local $N$-body
theory are presented in Section 6, while in Section 7 the $N$-body
Hamiltonian asymptotic approximation of the exact solution is developed
(THM.3). In Section 8 the validity of DGF is investigated, in connection
with the non-local Hamiltonian structure determined here and the
corresponding instant-form representation of Poincar\`{e} generators implied
by it. The extension of DGF to non-local system is presented in Section 9,
by means of the introduction of the notion of non-local Poisson brackets.
Then, an overview of the basic features of the \textquotedblleft
no-interaction\textquotedblright\ theorem is given in Section 10. Explicit
counter-examples to the theorem are determined, which are provided by the
existence of standard Hamiltonian forms for locally-isolated 1-body systems
and globally-isolated $N$-body systems (THM.4). The failure of the
\textquotedblleft no-interaction\textquotedblright\ theorem and the
non-applicability of its statements to the non-local Hamiltonian $N$-body
system considered here are discussed in Section 11. Finally, Section 12
provides a closing summary of the main results, while the mathematical
details concerning the derivation of the $N$-body action integral for the
binary interaction are reported in the Appendix.

\bigskip

\section{$N$-body EM current density and self 4-potential}

For the sake of clarity, let us first briefly recall the key points of the
formulation presented in Papers I and II dealing with the definitions of
extended charged particle and of the corresponding 4-current and self
4-potential. We shall assume that each finite-size particle is characterized
by a positive constant rest mass $m_{0}^{\left( i\right) }$ and a
non-vanishing constant charge $q^{\left( i\right) }$, for $i=1,N$, both
distributed on the same support $\partial \Omega _{\left( i\right) }$
(particle boundary). More precisely, for the $i$-th particle, the mass and
charge distributions can be defined as follows. Assuming that initially in a
time interval $[-\infty ,t_{o}]$ the $i$-th particle is at rest with respect
to an inertial frame (particle rest-frame $\mathcal{R}_{o}$ where the
external forces acting on the particle vanish identically), we shall assume
that:

1) In this frame there exists a point, hereafter referred to as \textit{%
center of symmetry (COS)}, whose position 4-vector $r_{\left( i\right)
COS}^{\mu }\equiv (ct,\mathbf{r}_{o})$ spans the Minkowski space-time $%
\mathcal{M}^{4}\subseteq \mathbb{R}^{4}$ with metric tensor $\eta _{\mu \nu
}\equiv $diag$\left( +1-1-1-1\right) $. With respect to the COS the support $%
\partial \Omega _{\left( i\right) }$ is a stationary spherical surface of
radius $\sigma _{\left( i\right) }>0$ of equation $\left( \mathbf{r-r}%
_{o}\right) ^{2}=\sigma _{\left( i\right) }^{2}$.

2) The $i$-th particle is \emph{quasi-rigid}, i.e., its mass and charge
distributions are stationary and spherically-symmetric on $\partial \Omega
_{\left( i\right) }$ (concerning the physical requirements assuring the
condition of rigidity we refer to the related discussion in Paper I).

3) Mass and charge densities do not possess pure spatial rotations.
Therefore, introducing for each particle the Euler angles $\alpha (s_{\left(
i\right) })\equiv \left\{ \varphi ,\vartheta ,\psi \right\} _{(s_{\left(
i\right) })}$ which define its spatial orientation (see definitions in Ref.%
\cite{Nodvik1964}), the condition of \emph{vanishing spatial rotation} is
obtained imposing that $\alpha (s_{\left( i\right) })=const.$ holds
identically. As a consequence, only the translational motion of charged
particles need to be taken into account.

In addition, as stated before, hard collisions occurring between the
particles are considered ignorable. As a consequence, for all particles the
equations of motion (\ref{STANDARD HAMILTONIAN FORM}) are assumed to hold
identically, namely for all $s_{\left( i\right) }\in I\equiv
\mathbb{R}
$.

For each extended particle the covariant expressions for the corresponding
charge and mass current densities readily follow (see Paper I). In
particular, these can be expressed in integral form respectively as:%
\begin{equation}
j^{\left( i\right) \mu }(r)=\frac{q^{\left( i\right) }c}{4\pi \sigma
_{\left( i\right) }^{2}}\int_{-\infty }^{+\infty }dsu^{\left( i\right) \mu
}(s)\delta (\left\vert x_{\left( i\right) }\right\vert -\sigma _{\left(
i\right) })\delta (s-s_{1\left( i\right) }),  \label{charge-current-i}
\end{equation}%
\begin{equation}
j_{mass}^{\left( i\right) \mu }(r)=\frac{m_{o}^{\left( i\right) }c}{4\pi
\sigma _{\left( i\right) }^{2}}\int_{-\infty }^{+\infty }dsu^{\left(
i\right) \mu }(s)\delta (\left\vert x_{\left( i\right) }\right\vert -\sigma
_{\left( i\right) })\delta (s-s_{1\left( i\right) }),  \label{mass_current-i}
\end{equation}%
where by definition $s_{1\left( i\right) }$ is the root of the algebraic
equation%
\begin{equation}
u_{\mu }^{\left( i\right) }(s_{1\left( i\right) })\left[ r^{\mu }-r^{\left(
i\right) \mu }\left( s_{1\left( i\right) }\right) \right] =0.
\end{equation}%
Here the notations are analogous to those given in Paper I, so that in
particular $u^{\left( i\right) \mu }(s_{\left( i\right) })\equiv \frac{%
dr^{\left( i\right) \mu }\left( s_{\left( i\right) }\right) }{ds_{\left(
i\right) }}$ is the 4-velocity of the COS for the $i$-th particle, while%
\begin{equation}
x_{\left( i\right) }^{\mu }=r^{\mu }-r^{\left( i\right) \mu }\left(
s_{\left( i\right) }\right) .
\end{equation}

Finally, following the equivalent derivations given in Papers I and II, the
non-divergent EM self 4-potential $A_{\mu }^{(self)\left( i\right) }(r)$ for
the single (namely, $i$-th) extended particle can be readily obtained as
well. It is sufficient to report here the solution for $A_{\mu
}^{(self)\left( i\right) }(r)$ which is valid in the external domain with
respect to the spherical shell of the same particle. For a generic
displacement 4-vector $X^{\left( i\right) \mu }\in M^{4}$ of the form%
\begin{equation}
X^{\left( i\right) \mu }=r^{\mu }-r^{\left( i\right) \mu }\left( s_{\left(
i\right) }\right) ,  \label{aaa}
\end{equation}%
which is subject to the constraint%
\begin{equation}
X^{\left( i\right) \mu }u_{\mu }^{\left( i\right) }(s_{\left( i\right) })=0,
\label{bbb}
\end{equation}%
this sub-domain is defined by the inequality%
\begin{equation}
X^{^{\left( i\right) }\mu }X_{\mu }^{\left( i\right) }\leq -\sigma _{\left(
i\right) }^{2}\emph{.}  \label{extdom}
\end{equation}%
In such a set, $A_{\mu }^{(self)\left( i\right) }(r)$ is expressed in
integral form by the equation%
\begin{equation}
A_{\mu }^{(self)\left( i\right) }(r)=2q^{\left( i\right)
}\int_{1}^{2}dr_{\mu }^{\prime }\delta (\widehat{R}^{\left( i\right) \alpha }%
\widehat{R}_{\alpha }^{\left( i\right) }),  \label{self-i}
\end{equation}%
where $\widehat{R}^{\left( i\right) \alpha }$ is the bi-vector%
\begin{equation}
\widehat{R}^{\left( i\right) \alpha }=r^{\alpha }-r^{\left( i\right) \prime
\alpha },
\end{equation}%
with $r^{\left( i\right) \prime \alpha }\equiv r^{\left( i\right) \prime
\alpha }(s_{\left( i\right) }^{\prime })$ being the $i$-th particle COS
4-vector evaluated at the retarded proper time $s_{\left( i\right) }^{\prime
}$, obtained as the causal root of the equation $\widehat{R}^{\left(
i\right) \alpha }\widehat{R}_{\alpha }^{\left( i\right) }=0$. As remarked in
Paper II, Eq.(\ref{self-i}) formally coincides with the analogous solution
holding for a point particle. However, in difference with the point-particle
case, $A_{\mu }^{(self)\left( i\right) }(r)$ can be defined everywhere\ in $%
M^{4}$ in such a way to be non-divergent (see Paper I).

\bigskip

\section{The non-local $N$-body action integral}

In this section we formulate the $N$-body Hamilton action functional
suitable for the variational treatment of a system of $N$ finite-size
charged particles subject to external, binary and self EM interactions. In
such a case, in analogy to Paper II, the action integral can be conveniently
expressed in hybrid super-abundant variables as follows:%
\begin{equation}
S_{N}(r,u,\left[ r\right] )=\sum_{i=1,N}\left[ S_{M}^{\left( i\right)
}(r,u)+S_{C}^{\left( ext\right) \left( i\right) }(r)+S_{C}^{\left(
self\right) \left( i\right) }(r,\left[ r\right] )+S_{C}^{(bin)\left(
i\right) }(r,\left[ r\right] )\right]  \label{pointp}
\end{equation}%
(\emph{non-local action integral}). As in Papers I and II, $r$ and $u$
represent \textit{local} dependences with respect to the 4-vector position
and the 4-velocity, while $\left[ r\right] $ stands for \textit{non-local}
dependences with respect to the 4-vector position. In particular, the latter
are included only via the functionals produced by the EM-coupling with the
self and binary EM fields for the $i$-th particle, namely $S_{C}^{\left(
self\right) \left( i\right) }$ and $S_{C}^{(bin)\left( i\right) }$. Instead,
$S_{M}^{\left( i\right) }$ and $S_{C}^{\left( ext\right) \left( i\right) }$
identify for each particle the functionals produced by the inertial mass and
by the EM-coupling with the external EM field. We stress that the
functionals $S_{M}^{\left( i\right) }(r,u)$, $S_{C}^{\left( ext\right)
\left( i\right) }(r)$ and $S_{C}^{\left( self\right) \left( i\right) }(r,%
\left[ r\right] )$ are formally analogous to the case of a 1-body problem
treated in Papers I and II and can be represented as line-integrals (see
below). We now proceed evaluating explicitly the new contribution $%
S_{C}^{(bin)\left( i\right) }(r,\left[ r\right] )$.

\bigskip

\subsection{$S_{C}^{\left( bin\right) \left( i\right) }(r,\left[ r\right] )$%
: EM coupling with the binary-interaction field}

The action integral $S_{C}^{\left( bin\right) \left( i\right) }(r,\left[ r%
\right] )$ containing the coupling between the EM field generated by
particle $j$, for $j=1,N$, and the electric 4-current of particle $i$ is of
critical importance. Its evaluation is similar to that of the action
integral of the self-interaction outlined in Paper I. For the sake of
clarity, in this subsection we present the relevant results, while the
details of the mathematical derivation are reported in the Appendix.
According to the standard approach \cite{LL}, $S_{C}^{\left( bin\right)
\left( i\right) }(r,\left[ r\right] )$ is defined as the 4-scalar%
\begin{equation}
S_{C}^{(bin)\left( i\right) }(r,\left[ r\right] )=\sum_{\substack{ j=1,N  \\ %
i\neq j}}S_{C}^{(bin)\left( ij\right) }(r,\left[ r\right] ),
\label{N-body-funct}
\end{equation}%
where $S_{C}^{(bin)\left( ij\right) }(r,\left[ r\right] )$ is defined as%
\begin{equation}
S_{C}^{(bin)\left( ij\right) }(r,\left[ r\right] )=\int_{1}^{2}d\Omega \frac{%
1}{c^{2}}A^{(self)\left( i\right) \mu }(r)j_{\mu }^{\left( j\right) }\left(
r\right) ,  \label{act1ij}
\end{equation}%
with $A^{(self)\left( i\right) \mu }(r)$ being the EM 4-potential generated
by particle $i$ at 4-position $r$, whose expression is given by Eq.(\ref%
{self-i}). In addition, $j_{\mu }^{\left( j\right) }\left( r\right) $ is the
4-current carried by particle $j$ evaluated at the same 4-position and given
by Eq.(\ref{charge-current-i}), while $d\Omega $ is the invariant 4-volume
element. In particular, in an inertial frame $S_{I}$ with Minkowski metric
tensor $\eta _{\mu \nu }$, this can be represented as $d\Omega =cdtdxdydz,$
where $\left( x,y,z\right) $ are orthogonal Cartesian coordinates. As shown
in the Appendix, an explicit evaluation of the action integral (\ref{act1ij}%
) yields the following representation:%
\begin{equation}
S_{C}^{(bin)\left( ij\right) }(r,\left[ r\right] )=\frac{2q^{\left( i\right)
}q^{\left( j\right) }}{c}\int_{1}^{2}dr_{\mu }^{\left( i\right) }\left(
s_{\left( i\right) }\right) \int_{1}^{2}dr^{\left( j\right) \mu }(s_{\left(
j\right) })\delta (\widetilde{R}^{\left( ij\right) \alpha }\widetilde{R}%
_{\alpha }^{\left( ij\right) }-\sigma _{\left( j\right) }^{2}),
\label{act1ij-bis}
\end{equation}%
where $s_{\left( i\right) }$ and $s_{\left( j\right) }$ are respectively the
proper times of particles $i$ and $j$, while $\widetilde{R}^{\left(
ij\right) \alpha }$ denotes%
\begin{equation}
\widetilde{R}^{\left( ij\right) \alpha }\equiv r^{\left( j\right) \alpha
}\left( s_{\left( j\right) }\right) -r^{\left( i\right) \alpha }(s_{\left(
i\right) }).
\end{equation}%
It is worth pointing out the following basic properties of the functional $%
S_{C}^{(bin,i)\left( ij\right) }$. First, it is a non-local functional in
the sense that it contains a coupling between the \textquotedblleft
past\textquotedblright\ and the \textquotedblleft future\textquotedblright\
of the particles of the $N$-body system. In fact it can be equivalently
represented as%
\begin{equation}
S_{C}^{(bin)\left( ij\right) }(r,\left[ r\right] )=\frac{2q^{\left( i\right)
}q^{\left( j\right) }}{c}\int_{-\infty }^{+\infty }ds_{\left( i\right) }%
\frac{dr_{\mu }^{\left( i\right) }\left( s_{\left( i\right) }\right) }{%
ds_{\left( i\right) }}\int_{-\infty }^{+\infty }ds_{\left( j\right) }\frac{%
dr^{\mu }(s_{\left( j\right) })}{ds_{\left( j\right) }}\delta (\widetilde{R}%
^{\left( ij\right) \alpha }\widetilde{R}_{\alpha }^{\left( ij\right)
}-\sigma _{\left( j\right) }^{2}).  \label{bin-ij}
\end{equation}%
Furthermore, the $N$-body system functional (\ref{N-body-funct}) is
symmetric, namely it fulfills the property%
\begin{equation}
\sum_{i,j=1,N}S_{C}^{(bin)\left( ij\right) }(r_{A},\left[ r_{B}\right]
)=\sum_{i,j=1,N}S_{C}^{(bin)\left( ji\right) }(r_{B},\left[ r_{A}\right] ),
\end{equation}%
where $r_{A}$ and $r_{B}$ are two arbitrary curves of the $N$-body system.

\subsection{The non-local $N$-body variational Lagrangian}

In this section we provide a line-integral representation of the \textit{%
Hamilton functional} $S_{N}$ in the form%
\begin{equation}
S_{N}=\sum_{i=1,N}\int_{-\infty }^{+\infty }ds_{\left( i\right)
}L_{1}^{\left( i\right) }(r,u,\left[ r\right] )\equiv
\sum_{i=1,N}\int_{-\infty }^{+\infty }\Upsilon _{\left( i\right) }(r,u,\left[
r\right] ),  \label{Ham-funct}
\end{equation}%
where $\Upsilon _{\left( i\right) }(r,\left[ r\right] ,u)$ and $%
L_{1}^{\left( i\right) }(r,\left[ r\right] ,u)$ are respectively the $i$-th
particle non-local contributions to the fundamental Lagrangian differential
form and to the corresponding \emph{non-local variational\ Lagrangian}.
Invoking Eq.(\ref{bin-ij}) and recalling also the results of Paper II, $%
L_{1}^{\left( i\right) }(r,\left[ r\right] ,u)$ can be written as%
\begin{equation}
L_{1}^{\left( i\right) }(r,u,\left[ r\right] )=L_{M}^{\left( i\right)
}(r,u)+L_{C}^{(ext)\left( i\right) }(r)+L_{C}^{(self)\left( i\right) }(r,%
\left[ r\right] )+L_{C}^{(bin)\left( i\right) }(r,\left[ r\right] ),
\label{Lagr-ij}
\end{equation}%
where $L_{M}^{\left( i\right) }(r,u),L_{C}^{(ext)\left( i\right) }(r)$ and $%
L_{C}^{(self)\left( i\right) }(r,\left[ r\right] ),L_{C}^{(bin)\left(
i\right) }(r,\left[ r\right] )$ denote respectively the local and non-local
terms. In particular, the first one is the contribution carried by the
inertial term, while $L_{C}^{(ext)\left( i\right) }$, $L_{C}^{(self)\left(
i\right) }$ and $L_{C}^{(bin)\left( i\right) }$ identify respectively the
external, self and binary EM-field-coupling Lagrangians. These are defined
as follows:%
\begin{eqnarray}
L_{M}^{\left( i\right) }(r,u) &\equiv &m_{o}^{\left( i\right) }cu_{\mu
}^{\left( i\right) }\left[ \frac{dr^{\left( i\right) \mu }}{ds_{\left(
i\right) }}-\frac{1}{2}u^{\left( i\right) \mu }\right] ,  \label{Lmass-ij} \\
L_{C}^{(ext)\left( i\right) }(r) &\equiv &\frac{q^{\left( i\right) }}{c}%
\frac{dr^{\left( i\right) \mu }}{ds_{\left( i\right) }}\overline{A}_{\mu
}^{(ext)\left( i\right) }(r^{\left( i\right) }(s_{\left( i\right) }),\sigma
_{\left( i\right) }),  \label{Lext-ij} \\
L_{C}^{(self)\left( i\right) }(r,\left[ r\right] ) &\equiv &\frac{q^{\left(
i\right) }}{c}\frac{dr^{\left( i\right) \mu }}{ds_{\left( i\right) }}%
\overline{A}_{\mu }^{(self)\left( i\right) },  \label{Lself-ij} \\
L_{C}^{(bin)\left( i\right) }(r,\left[ r\right] ) &\equiv &\sum_{\substack{ %
j=1,N  \\ i\neq j}}L_{C}^{(bin)\left( ij\right) }(r,\left[ r\right] )=\frac{%
q^{\left( i\right) }}{c}\frac{dr^{\left( i\right) \mu }}{ds_{\left( i\right)
}}\sum_{\substack{ j=1,N  \\ i\neq j}}\overline{A}_{\mu }^{(bin)\left(
ij\right) }\left( \sigma _{\left( j\right) }\right) .  \label{Lbin-ij-1}
\end{eqnarray}%
Here, $\overline{A}_{\mu }^{(ext)\left( i\right) },$ $\overline{A}_{\mu
}^{(self)\left( i\right) }$ and $\overline{A}_{\mu }^{(bin)\left( ij\right)
} $ denote the surface-averages performed on the $i$-th particle boundary $%
\partial \Omega _{\left( i\right) }$ (see Paper I) respectively of the
external, self and binary EM 4-potentials. In particular, $\overline{A}_{\mu
}^{(self)\left( i\right) }$ and $\overline{A}_{\mu }^{(bin)\left( ij\right)
} $ are defined as%
\begin{eqnarray}
\overline{A}_{\mu }^{(self)\left( i\right) } &\equiv &2q^{\left( i\right)
}\int_{1}^{2}dr_{\mu }^{\left( i\right) \prime }\delta (\widetilde{R}%
^{\left( i\right) \mu }\widetilde{R}_{\mu }^{\left( i\right) }-\sigma
_{\left( i\right) }^{2}),  \label{Aself-i} \\
\overline{A}_{\mu }^{(bin)\left( ij\right) }\left( \sigma _{\left( j\right)
}\right) &\equiv &2q^{\left( j\right) }\int_{-\infty }^{+\infty }ds_{\left(
j\right) }\frac{dr^{\mu }(s_{\left( j\right) })}{ds_{\left( j\right) }}%
\delta (\widetilde{R}^{\left( ij\right) \alpha }\widetilde{R}_{\alpha
}^{\left( ij\right) }-\sigma _{\left( j\right) }^{2}).  \label{Aself-ij}
\end{eqnarray}%
In addition, $\widetilde{R}^{\left( i\right) \mu }$ is the bi-vector%
\begin{equation}
\widetilde{R}^{\left( i\right) \mu }\equiv r^{\left( i\right) \mu }\left(
s_{\left( i\right) }\right) -r^{\left( i\right) \mu }(s_{\left( i\right)
}^{\prime }),  \label{Rtilda}
\end{equation}%
with $s_{\left( i\right) }$ and $s_{\left( i\right) }^{\prime }$ denoting
respectively \textquotedblleft present\textquotedblright\ and
\textquotedblleft retarded\textquotedblright\ proper times of the $i$-th
particle.

$\bigskip $

\section{Non-local $N$-body variational principle and standard Lagrangian
form}

Let us now proceed constructing the explicit form of the $N$-body
relativistic equations of motion for each extended charged particle \textit{%
in the presence of EM interactions }(i.e., including \textit{external,
binary and self EM interactions}). This is achieved by adopting for the $N$%
-body problem a \textit{synchronous variational principle} \cite%
{Tessarotto2006,Tessarotto2008c} which, in analogy with the approach
developed in Papers I and II, can be expressed in terms of the
super-abundant hybrid (i.e., generally non-Lagrangian) variables%
\begin{equation}
\emph{\ }f^{\left( i\right) }(s_{\left( i\right) })\equiv \left[ r^{\left(
i\right) \mu }(s_{\left( i\right) }),u_{\mu }^{\left( i\right) }(s_{\left(
i\right) })\right] ,  \label{variational--functions}
\end{equation}%
and for a suitable functional class of variations $\left\{ f\right\} $. The
latter is identified with the set of real functions of class $C^{k}(\mathbb{R%
}),$ with $k\geq 2,$ and fixed endpoints which are prescribed for each
particle $i=1,N$ at suitable proper times $s_{\left( i\right) 1}$\ and $%
s_{\left( i\right) 2},$ with $s_{\left( i\right) 1}<$ $s_{\left( i\right) 2}$%
, i.e.,%
\begin{equation}
\left\{ f\right\} \equiv \left\{
\begin{array}{c}
\emph{\ }f^{\left( i\right) }(s_{\left( i\right) }):\emph{\ }f^{\left(
i\right) }(s_{\left( i\right) })\in C^{k}(\mathbb{R}); \\
\emph{\ }f^{\left( i\right) }(s_{\left( i\right) j})=f_{j}^{\left( i\right)
}; \\
i=1,N;\emph{\ }j=1,2\text{ \emph{and} }k\geq 2%
\end{array}%
\right\} .  \label{FUNCTIONAL CLASS}
\end{equation}%
It follows that by construction the variational derivatives of the Hamilton
functional $S_{N}$ (see Eq.(\ref{Ham-funct})) are performed in terms of
synchronous variations, i.e., by keeping constant the $i$-th particle proper
time $s_{\left( i\right) }$. The result is expressed by the following
theorem.

\bigskip

\textbf{THM.1 - }$N$\textbf{-body hybrid synchronous Hamilton variational
principle}

\emph{Given validity of the prerequisites \#1-\#5 for the }$N$\emph{-body
system, let us assume that:}

\begin{enumerate}
\item \emph{The Hamilton action }$S_{N}(r,u,\left[ r\right] )$\emph{\ is
defined by Eq.(\ref{Ham-funct}).}

\item \emph{The real functions }$f^{\left( i\right) }(s_{\left( i\right) })$
\emph{in the functional class }$\left\{ f\right\} $ \emph{[see Eq.(\ref%
{FUNCTIONAL CLASS})]}\emph{\ are identified with the super-abundant variables%
} \emph{(\ref{variational--functions}) which are subject to synchronous
variations }$\delta f^{\left( i\right) }(s_{\left( i\right) })\equiv $\emph{%
\ }$f^{\left( i\right) }(s_{\left( i\right) })-$\emph{\ }$f_{1}^{\left(
i\right) }(s_{\left( i\right) }).$\emph{\ The latter belong to the
functional class of synchronous variations }$\left\{ \delta f^{\left(
i\right) }\right\} $\emph{, with}%
\begin{equation}
\delta f_{k}^{\left( i\right) }(s_{\left( i\right) })\emph{\ }=f_{k}^{\left(
i\right) }(s_{\left( i\right) })-f_{1k}^{\left( i\right) }(s_{\left(
i\right) }),  \label{SYNCHRONOUS FUNCTIONAL
CLASS}
\end{equation}%
\emph{for }$k=1,2,$ $\forall f^{\left( i\right) }(s_{\left( i\right)
}),f_{1}^{\left( i\right) }(s_{\left( i\right) })\in \left\{ f\right\} $%
\emph{.}

\item \emph{The extremal curves }$f^{\left( i\right) }(s_{\left( i\right)
})\in \left\{ f\right\} $ \emph{for }$S_{N}$\emph{, which are solutions of
the E-L equations}%
\begin{equation}
\frac{\delta S_{N}(r,u,\left[ r\right] )}{\delta f^{\left( i\right)
}(s_{\left( i\right) })}=0,  \label{Hamilton-princ}
\end{equation}%
\emph{exist for arbitrary variations} $\delta f^{\left( i\right) }(s_{\left(
i\right) })$ \emph{(hybrid synchronous Hamilton variational principle).}

\item \emph{If the curves }$r^{\left( i\right) \mu }(s_{\left( i\right) })$%
\emph{, for} $i=1,N$ \emph{are all extremal, each line element }$ds_{\left(
i\right) }$\emph{\ satisfies the constraint}%
\begin{equation}
ds_{\left( i\right) }^{2}=\eta _{\mu \nu }dr^{\left( i\right) \mu
}(s_{\left( i\right) })dr^{\left( i\right) \nu }(s_{\left( i\right) })\emph{.%
}  \label{constraint-aa}
\end{equation}

\item \emph{The 4-vector field }$A_{\mu }^{(ext)}(r)$ \emph{is suitably
smooth in the whole Minkowski space-time }$M^{4}$\emph{.}

\item \emph{The E-L equations for the extremal curves }$r^{\left( i\right)
\mu }(s_{\left( i\right) })$\emph{\ are determined consistently with the
Einstein causality principle.}

\item \emph{All the synchronous variations }$\delta f_{k}^{\left( i\right)
}(s_{\left( i\right) })$ \emph{(}$\emph{k=1,2}$ \emph{and }$i=1,N$\emph{)
are considered as being independent.}
\end{enumerate}

\emph{It then follows that the E-L equations for $u^{\left( i\right) \mu }$
and }$r^{\left( i\right) \mu }$ \emph{following from the synchronous hybrid
Hamilton variational principle (\ref{Hamilton-princ})} \emph{give
respectively}%
\begin{eqnarray}
&&\left. \frac{\delta S_{N}}{\delta u_{\mu }^{\left( i\right) }}%
=m_{o}^{\left( i\right) }cdr^{\left( i\right) \mu }-m_{o}^{\left( i\right)
}cu^{\left( i\right) \mu }ds_{\left( i\right) }=0,\right.  \label{Eq.1RR} \\
&&\left. \frac{\delta S_{N}}{\delta r^{\left( i\right) \mu }(s_{\left(
i\right) })}=-m_{o}^{\left( i\right) }cdu_{\mu }^{\left( i\right)
}(s_{\left( i\right) })+\frac{q^{\left( i\right) }}{c}F_{\mu \nu
}^{(tot)\left( i\right) }dr^{\left( i\right) \nu }(s_{i})=0,\right.
\label{Eq.2RR}
\end{eqnarray}%
\emph{where }$F_{\mu \nu }^{(tot)\left( i\right) }$ \emph{is the total
Faraday tensor acting on particle }$i$ \emph{and given by}%
\begin{equation}
F_{\mu \nu }^{(tot)\left( i\right) }\equiv \overline{F}_{\mu \nu
}^{(ext)\left( i\right) }+\overline{F}_{\mu \nu }^{\left( self\right) \left(
i\right) }+\overline{F}_{\mu \nu }^{\left( bin\right) \left( i\right) },
\label{Fmunu-1}
\end{equation}%
\emph{where all quantities are intended as surface-averages on the }$i$\emph{%
-th particle shell-surface }$\partial \Omega _{\left( i\right) }$\emph{.
Eqs.(\ref{Eq.1RR}) and (\ref{Eq.2RR}) are hereon referred to as }$N$\emph{%
-body equations of motion. In particular:}

\emph{1) }$\overline{F}_{\mu \nu }^{(ext)\left( i\right) }\left( r^{\left(
i\right) }\right) \equiv \partial _{\mu }\overline{A}_{\nu
}^{(ext)}-\partial _{\nu }\overline{A}_{\mu }^{(ext)}$\emph{\ is the
antisymmetric Faraday tensor of the external EM field evaluated on the
extremal curve }$r^{\left( i\right) \mu }=r^{\left( i\right) \mu }\left(
s_{\left( i\right) }\right) .$

\emph{2) }$\overline{F}_{\mu \nu }^{\left( self\right) \left( i\right)
}\left( r^{\left( i\right) },\left[ r^{\left( i\right) }\right] \right)
\equiv \overline{F}_{\mu \nu }^{\left( self\right) \left( i\right) }\left(
r^{\left( i\right) }\left( s_{\left( i\right) }\right) ,r^{\left( i\right)
}\left( s_{\left( i\right) }^{\prime }\right) \right) $ \emph{is the
non-local antisymmetric Faraday tensor produced by the EM self-field of the }%
$i$\emph{-th particle and acting on the same particle. This is given by}%
\begin{equation}
\overline{F}_{\mu \nu }^{\left( self\right) \left( i\right) }=2\left[
\partial _{\mu }\overline{A}_{\nu }^{(self)}-\partial _{\nu }\overline{A}%
_{\mu }^{(self)}\right] ,
\end{equation}%
\emph{namely}%
\begin{equation}
\overline{F}_{\mu \nu }^{\left( self\right) \left( i\right) }=-\left[ \frac{%
2q^{\left( i\right) }}{\left\vert \widetilde{R}^{\left( i\right) \alpha
}u_{\alpha }^{\left( i\right) }(s_{\left( i\right) }^{\prime })\right\vert }%
\frac{d}{ds_{\left( i\right) }^{\prime }}\left\{ \frac{u_{\mu }^{\left(
i\right) }(s_{\left( i\right) }^{\prime })\widetilde{R}_{\nu }^{\left(
i\right) }-u_{\nu }^{\left( i\right) }(s_{\left( i\right) }^{\prime })%
\widetilde{R}_{\mu }^{\left( i\right) }}{\widetilde{R}^{\left( i\right)
\alpha }u_{\alpha }^{\left( i\right) }(s_{i}^{\prime })}\right\} \right]
_{s_{\left( i\right) }^{\prime }=s_{\left( i\right) }-s_{\left( i\right)
ret}},  \label{COVARIANT FORM}
\end{equation}%
\emph{where the }$\emph{delay-time}$\emph{\ }$s_{\left( i\right) ret}$\emph{%
\ is the positive (causal) root of the 1-particle delay-time equation}%
\begin{equation}
\widetilde{R}^{\left( i\right) \alpha }\widetilde{R}_{\alpha }^{\left(
i\right) }-\sigma _{\left( i\right) }^{2}=0.  \label{delayshell}
\end{equation}

\emph{3) }$\overline{F}_{\mu \nu }^{\left( bin\right) \left( i\right) }$%
\emph{\ is the non-local antisymmetric Faraday tensor produced on particle }$%
i$\emph{\ by the action of all the remaining particles, i.e. }$\overline{F}%
_{\mu \nu }^{\left( bin\right) \left( i\right) }\equiv \sum_{\substack{ %
j=1,N  \\ i\neq j}}\overline{F}_{\mu \nu }^{\left( bin\right) \left(
ij\right) }\left( r^{\left( i\right) },\left[ r^{\left( j\right) }\right]
,\sigma _{\left( i\right) },\sigma _{\left( j\right) }\right) $\emph{, where}%
\begin{equation}
\overline{F}_{\mu \nu }^{\left( bin\right) \left( ij\right) }\left(
r^{\left( i\right) },\left[ r^{\left( j\right) }\right] ,\sigma _{\left(
i\right) }\text{\emph{,}}\sigma _{\left( j\right) }\right) =\left[ H_{\mu
\nu }^{\left( ij\right) }\left( s_{\left( i\right) },s_{\left( j\right)
}\right) \right] _{s_{\left( j\right) }=s_{\left( i\right) }^{\left(
A\right) }\left( \sigma _{\left( i\right) }\right) }+\left[ H_{\mu \nu
}^{\left( ij\right) }\left( s_{\left( i\right) },s_{\left( j\right) }\right) %
\right] _{s_{\left( j\right) }=s_{\left( ij\right) }^{\left( B\right)
}\left( \sigma _{\left( j\right) }\right) }.  \label{Fbin-ij}
\end{equation}%
\emph{Here the notation is as follows. }$H_{\mu \nu }^{\left( ij\right) }$%
\emph{\ is defined as}%
\begin{equation}
H_{\mu \nu }^{\left( ij\right) }\left( s_{\left( i\right) },s_{\left(
j\right) }\right) =-\frac{q^{\left( j\right) }}{\left\vert \widetilde{R}%
^{\left( ij\right) \alpha }u_{\alpha }^{\left( j\right) }(s_{\left( j\right)
})\right\vert }\frac{d}{ds_{\left( j\right) }}\left\{ \frac{u_{\mu }^{\left(
j\right) }(s_{\left( j\right) })\widetilde{R}_{\nu }^{\left( ij\right)
}-u_{\nu }^{\left( i\right) }(s_{\left( j\right) })\widetilde{R}_{\mu
}^{\left( ij\right) }}{\widetilde{R}^{\left( ij\right) \alpha }u_{\alpha
}^{\left( j\right) }(s_{\left( j\right) })}\right\} ,  \label{Fbin-ij-2}
\end{equation}%
\emph{while the delay-time }$s_{\left( j\right) }=s_{\left( i\right)
}^{\left( A\right) }\left( \sigma _{\left( i\right) }\right) $\emph{\ and }$%
s_{\left( j\right) }=s_{\left( ij\right) }^{\left( B\right) }\left( \sigma
_{\left( j\right) }\right) $\emph{\ are respectively the positive (causal)
roots of the 2-particle delay-time equations}%
\begin{eqnarray}
\widetilde{R}^{\left( i\right) \alpha }\widetilde{R}_{\alpha }^{\left(
i\right) }-\sigma _{\left( i\right) }^{2} &=&0,  \label{delay-ij-1} \\
\widetilde{R}^{\left( ij\right) \alpha }\widetilde{R}_{\alpha }^{\left(
ij\right) }-\sigma _{\left( j\right) }^{2} &=&0.  \label{delay-ij-2}
\end{eqnarray}%
\emph{Therefore, }$s_{\left( i\right) }^{\left( A\right) }$\emph{\ and }$%
s_{\left( ij\right) }^{\left( B\right) }$ \emph{depend respectively on }$%
\sigma _{\left( i\right) }$ \emph{and }$\sigma _{\left( j\right) }$\emph{.}

\emph{Proof} - The proof is analogous to the corresponding 1-body problem
detailed in THM.1 of Paper 1. Indeed, since the Dirac-deltas $\delta (%
\widetilde{R}^{\left( i\right) \mu }\widetilde{R}_{\mu }^{\left( i\right)
}-\sigma _{\left( i\right) }^{2})$ and $\delta (\widetilde{R}^{\left(
ij\right) \alpha }\widetilde{R}_{\alpha }^{\left( ij\right) }-\sigma
_{\left( j\right) }^{2})$ are independent of particle 4-velocities, the
variations with respect to $u_{\mu }^{\left( i\right) }$ deliver necessarily
the E-L equation (\ref{Eq.1RR}). To prove also Eq.(\ref{Eq.2RR}), we notice
that the synchronous variations of\ the functionals $S_{M}^{\left( i\right)
}(r,u),$ $S_{C}^{\left( ext\right) \left( i\right) }(r)$ and $S_{C}^{\left(
self\right) \left( i\right) }(r,\left[ r\right] )$ necessarily coincide with
those of the 1-body problem (see Papers I and II). Therefore, it is
sufficient to inspect the variational derivative of the non-local
binary-interaction functional $S_{C}^{(bin)\left( i\right) }(r,\left[ r%
\right] )$. Its variation with respect to $\delta r^{\left( i\right) \mu
}(s_{\left( i\right) })$ takes the form
\begin{equation}
\delta S_{C}^{(bin)\left( i\right) }=\sum_{\substack{ j=1,N  \\ i\neq j}}%
\left\{ \left[ \delta A+\delta B\right] _{\left( ij\right) }+\left[ \delta
A+\delta B\right] _{\left( ji\right) }\right\} ,  \label{Eq.---A.3}
\end{equation}%
where%
\begin{equation}
\begin{array}{c}
\delta A_{\left( ij\right) }\equiv -\frac{2q^{\left( i\right) }q^{\left(
j\right) }}{c}\eta _{\mu \nu }\int_{1}^{2}\delta r^{\left( i\right) \mu }d%
\left[ \int_{1}^{2}dr^{\left( j\right) \nu }\delta (\widetilde{R}^{\left(
ij\right) \alpha }\widetilde{R}_{\alpha }^{\left( ij\right) }-\sigma
_{\left( i\right) }^{2})\right] , \\
\delta B_{\left( ij\right) }\equiv \frac{2q^{\left( i\right) }q^{\left(
j\right) }}{c}\eta _{\alpha \beta }\int_{1}^{2}dr^{\left( j\right) \beta
}\int_{1}^{2}dr^{\left( i\right) \alpha }\delta r^{\left( i\right) \mu }%
\frac{\partial }{\partial r^{\left( i\right) \mu }}\delta (\widetilde{R}%
^{\left( ij\right) \alpha }\widetilde{R}_{\alpha }^{\left( ij\right)
}-\sigma _{\left( i\right) }^{2}),%
\end{array}
\label{EQ.---A.4}
\end{equation}%
and the second term $\left[ \delta A+\delta B\right] _{\left( ji\right) }$
follows by exchanging the particle indices. Then, using the chain rule and
integrating by parts, after elementary algebra Eqs.(\ref{Fbin-ij}) and (\ref%
{Fbin-ij-2}) follow (for details see Appendix A in Paper I). In agreement
with the Einstein causality principle the positive roots of the delay-time
equations (\ref{delay-ij-1}) and (\ref{delay-ij-2}) must be selected.

\textbf{Q.E.D.}

\bigskip

A few comments are here in order regarding the implications of THM.1.

\begin{enumerate}
\item \textbf{Coordinate-time parametrization of the N-body equations of
motion}

It is important to stress that for each $i$-th particle, its equations of
motion [in particular the E-L Eqs.(\ref{Eq.1RR}) and (\ref{Eq.2RR})] can be
parametrized in terms of the single coordinate time $t$ rather than the
corresponding particle proper time $s_{(i)}.$ This is obtained introducing
the representations in terms of the single coordinate (i.e., Laboratory) time%
$\ t\in I\in
\mathbb{R}
,$ namely letting for all $i=1,N$%
\begin{eqnarray}
r^{\left( i\right) \mu }(t) &\equiv &(ct,\mathbf{r}^{(i)}),  \notag \\
ds_{(i)} &=&\frac{cdt}{\gamma _{(i)}},
\label{COORDINATE-TIME PARAMETRIZATION}
\end{eqnarray}%
with $\gamma _{(i)}$ and $\mathbf{\beta }_{(i)}$ denoting the usual
relativistic factors%
\begin{eqnarray}
\gamma _{(i)} &=&\left( 1-\beta _{(i)}^{2}\right) ^{-1/2},  \label{GAMMA_i}
\\
\mathbf{\beta }_{(i)} &=&\mathbf{v}^{(i)}/c.  \label{BETA_i}
\end{eqnarray}%
This implies also that the 4-velocity can be represented as $u^{\left(
i\right) \mu }\equiv \frac{1}{c}\gamma _{(i)}^{-1}v^{\left( i\right) \mu }$
with $v^{\left( i\right) \mu }\equiv (c,\mathbf{v}^{(i)}).$ Hence, equations
(\ref{Eq.1RR}) and (\ref{Eq.2RR}) become respectively
\begin{eqnarray}
&&\left. m_{o}^{\left( i\right) }cdr^{\left( i\right) \mu }-m_{o}^{\left(
i\right) }cv^{\left( i\right) \mu }dt=0,\right.  \label{DIFFERENCE0EQ-1} \\
&&\left. -m_{o}^{\left( i\right) }cd\left[ \frac{\gamma _{(i)}}{c}v_{\mu
}^{\left( i\right) }(s_{\left( i\right) })\right] +\frac{q^{\left( i\right) }%
}{c}F_{\mu \nu }^{(tot)\left( i\right) }dr^{\left( i\right) \nu
}(s_{i})=0.\right.  \label{DIFFERENCE0EQ-2}
\end{eqnarray}

\item \textbf{Delay-time effects}

Delay-time effects which appear both in the EM RR and binary interactions
are due to the extended size of the charged particles. In particular, the
delay-time characterizing the self-interaction acting on particle $i$
depends only on the radius of the charge distribution of the same particle.
Instead, the delay-time appearing in the binary interaction experienced by
particle $i$ depends either on the radius $\sigma _{\left( i\right) }$ of
particle $i$ or on the radii $\sigma _{\left( j\right) }$ of all the
remaining particles. In the case of $N$-body system of like particles, such
that $\sigma _{\left( i\right) }=\sigma _{\left( j\right) }=\sigma $, the
two terms on the r.h.s. of Eq.(\ref{Eq.---A.3}) coincide yielding a single
delay-time contribution in Eq.(\ref{Fbin-ij}). Explicit evaluation of delay
times involves the construction of the positive (causal) roots of the
equations (\ref{delay-ij-1}) and (\ref{delay-ij-2}). Based on the
coordinate-time parametrization (\ref{COORDINATE-TIME PARAMETRIZATION}),\
these can be solved explicitly for the \emph{coordinate delay-time }$%
t_{(i)ret}\equiv t_{(i)}^{\prime }-t.$ The causal roots are in the two cases
respectively%
\begin{equation}
t_{(i)ret}(t)=\frac{1}{c}\sqrt{\left[ \mathbf{r}^{(i)}(t)\mathbf{-r}%
^{(i)}(t-t_{(i)ret}(t))\right] ^{2}+\sigma _{(i)}^{2}}>0,
\label{delay-coo-time-1}
\end{equation}%
\begin{equation}
t_{(ij)ret}(t)=\frac{1}{c}\sqrt{\left[ \mathbf{r}^{(i)}(t)\mathbf{-r}%
^{(j)}(t-t_{(ij)ret}(t))\right] ^{2}+\sigma _{(j)}^{2}}>0.
\label{delay-coo-time-2}
\end{equation}%
Notice that the same roots can also be equivalently represented in terms of
the corresponding particle proper times ($s_{(i)}$ for $i=1,N$). For this
purpose it is sufficient to introduce for the $i$-th particle proper time
the parametrization $s_{(i)}\equiv s_{(i)}(t)$ which is determined in terms
of the coordinate time $t$ by means of the equations (\ref{constraint-aa}).
It follows, in particular, that the proper delay-times corresponding to (\ref%
{delay-coo-time-1}) and (\ref{delay-coo-time-2}) become respectively $%
s_{(i)ret}(s_{(i)})\equiv s(t_{(i)ret}(t))$ and $s_{(ij)ret}(s_{(j)})\equiv
s(t_{(ij)ret}(t))$.
\end{enumerate}

\bigskip

Along the lines of the approach given in Paper II, it is immediate to show
that the hybrid-variable variational principle given in THM.1 can be given
an equivalent Lagrangian formulation. An elementary consequence is provided
by the following proposition.

\bigskip

\textbf{Corollary to THM.1 - Standard Lagrangian form of the }$N$\textbf{%
-body equations of motion}

\emph{Given validity of THM.1, let us introduce the non-local real function}%
\begin{equation}
L_{eff,N}=\sum_{i=1,N}L_{eff}^{\left( i\right) }(r,u,\left[ r\right] ),
\label{Leff-N}
\end{equation}%
\emph{where }$L_{eff,N}$\emph{\ is denoted as }$N$\emph{-body effective
Lagrangian, while }$L_{eff}^{\left( i\right) }(r,u,\left[ r\right] )$\emph{\
is defined as}%
\begin{equation}
L_{eff}^{\left( i\right) }(r,u,\left[ r\right] )\equiv L_{M}^{\left(
i\right) }(r,u)+L_{C}^{(ext)\left( i\right) }(r)+2L_{C}^{(self)\left(
i\right) }(r,\left[ r\right] )+L_{eff}^{(bin)\left( i\right) }(r,\left[ r%
\right] ).  \label{Leff-i}
\end{equation}%
\emph{Here }$L_{M}^{\left( i\right) },$\emph{\ }$L_{C}^{(ext)\left( i\right)
}$\emph{\ and }$L_{C}^{(self)\left( i\right) }$\emph{\ coincide with the
variational Lagrangians defined above (see Eqs.(\ref{Lmass-ij})-(\ref%
{Lself-ij})), while }$L_{eff}^{(bin)\left( i\right) }$\emph{\ is given by}%
\begin{equation}
L_{eff}^{(bin)\left( i\right) }(r,\left[ r\right] )\equiv \sum_{\substack{ %
j=1,N  \\ i\neq j}}\frac{2q^{\left( i\right) }q^{\left( j\right) }}{c}\frac{%
dr_{\mu }^{\left( i\right) }\left( s_{\left( i\right) }\right) }{ds_{\left(
i\right) }}\int_{-\infty }^{+\infty }ds_{\left( j\right) }\frac{dr^{\mu
}(s_{\left( j\right) })}{ds_{\left( j\right) }}K^{\left( ij\right) },
\label{Leff-bin}
\end{equation}%
\emph{with }$K^{\left( ij\right) }$\emph{\ being the sum of Dirac-deltas}%
\begin{equation}
K^{\left( ij\right) }\equiv \delta (\widetilde{R}^{\left( ij\right) \alpha }%
\widetilde{R}_{\alpha }^{\left( ij\right) }-\sigma _{\left( j\right)
}^{2})+\delta (\widetilde{R}^{\left( ij\right) \alpha }\widetilde{R}_{\alpha
}^{\left( ij\right) }-\sigma _{\left( i\right) }^{2}).  \label{Kij}
\end{equation}%
\emph{Then, the E-L equations (\ref{Eq.1RR}) and (\ref{Eq.2RR}) coincide
with the E-L equations in standard form (see Papers I and II) determined in
terms of the }$N$\emph{-body effective Lagrangian }$L_{eff,N}$\emph{. In
particular, the E-L equations in standard form for the }$i$\emph{-th
particle become}%
\begin{eqnarray}
&&\left. \frac{\partial L_{eff,N}}{\partial u_{\mu }^{\left( i\right)
}(s_{\left( i\right) })}=0,\right.  \label{EL-2} \\
&&\left. F_{\mu }^{\left( i\right) }(r)L_{eff,N}=0,\right.  \label{EL-3}
\end{eqnarray}%
\emph{where}%
\begin{equation}
F_{\mu }^{\left( i\right) }(r)\equiv \frac{d}{ds_{\left( i\right) }}\frac{%
\partial }{\partial \frac{dr^{\left( i\right) \mu }(s_{\left( i\right) })}{%
ds_{\left( i\right) }}}-\frac{\partial }{\partial r^{\left( i\right) \mu
}(s_{\left( i\right) })}
\end{equation}%
\emph{denotes the E-L differential operator.}

\emph{Proof} - The proof is based on THM.2 of Paper I and by noting that the
E-L differential operator acts only on local quantities. Hence, the
equivalence of Eqs.(\ref{EL-2})-(\ref{EL-3}) with the corresponding E-L
equations (\ref{Eq.1RR}) and (\ref{Eq.2RR}) follows by elementary algebra
and in view of the identity $F_{\mu }^{\left( i\right) }(r)L_{eff,N}=F_{\mu
}^{\left( i\right) }(r)L_{eff}^{\left( i\right) }$.

\textbf{Q.E.D.}

\bigskip

To conclude this Section a final remark is necessary regarding the
functional setting of the\textbf{\ }$N$-body equations of motion given above.

We first notice that the E-L equations (\ref{Eq.1RR}) and (\ref{Eq.2RR}),
and the equivalent Lagrange equations in standard form (\ref{EL-3}), imply
for all $i=1,N$ the second-order delay-type ODEs%
\begin{equation}
m_{o}^{\left( i\right) }c\frac{d^{2}r^{\left( i\right) \mu }(s_{\left(
i\right) })}{ds_{(i)}^{2}}=\frac{q^{\left( i\right) }}{c}F_{\nu
}^{(tot)\left( i\right) \mu }dr^{\left( i\right) \nu }(s_{i}).
\label{N-BODY-RR-EQUATION}
\end{equation}%
Let us introduce the Lagrangian state $\mathbf{w}\equiv \left( \left(
r^{(i)},u^{(i)}\right) ,i=1,N\right) ,$ with $u^{\left( i\right) \mu }\equiv
\frac{dr^{\left( i\right) \mu }}{ds_{(i)}}$, spanning the $N$-body
phase-space $\Gamma _{N}\equiv \prod\limits_{i=1,N}\Gamma _{1(i)},$ where $%
\Gamma _{1(i)}=M_{(i)}^{(4)}\times U_{(i)}^{(4)}$ and $U_{(i)}^{(4)}\equiv
\mathbb{R}
^{4}$ indicate respectively the corresponding $1$-body phase and velocity
spaces, the latter endowed with a metric tensor $\eta _{\mu \nu }$. To
define the initial conditions, let us make use of the coordinate-time
parametrization (\ref{COORDINATE-TIME PARAMETRIZATION}), denoting $\widehat{%
\mathbf{w}}(t)\equiv \left( \left( \widehat{r}^{(i)}(t),\widehat{u}%
^{(i)}(t)\right) ,i=1,N\right) $ and $\widehat{r}^{(i)}(t)\equiv
r^{(i)}(s_{(i)}(t)),$ $\widehat{u}^{(i)}(t)\equiv u^{(i)}(s_{(i)}(t)).$
Then, a well-posed problem for Eqs.(\ref{N-BODY-RR-EQUATION}) is obtained
prescribing the \emph{initial history set }$\left\{ \widehat{\mathbf{w}}%
\right\} _{t_{0}}\subset \Gamma _{N}$\emph{. }For an arbitrary \emph{%
coordinate initial time }$t_{0}\in I\equiv
\mathbb{R}
,$ this is defined as the ensemble of initial states
\begin{equation}
\left\{ \widehat{\mathbf{w}}\right\} _{t_{0}}\equiv \left\{ \left( \left(
\widehat{r}^{(i)}(t),\widehat{u}^{(i)}(t)\right) \in
C^{(k-1)}(I),i=1,N\right) ,\forall t\in \left[ t_{0}-t_{ret}^{\max
}(t_{0}),t_{0}\right] ,k\geq 2\right\} .  \label{INITIAL HISTORY SET}
\end{equation}%
Here, for a given initial (coordinate) time $t_{0}\in I,$ $t_{ret}^{\max
}(t_{0})$ denotes the maximum (for all particles) of the delay-times $%
t_{(i)ret}(t_{0})$ and $t_{(ij)ret}(t_{0}),$ namely
\begin{equation}
t_{ret}^{\max }(t_{0})=\max \left\{ t_{(i)ret}(t_{0}),t_{(ij(ret}(t_{0}),%
\text{ }\forall i,j=1,N\right\} .  \label{MAX-DELAY TIME}
\end{equation}%
Solutions of Eqs.(\ref{N-BODY-RR-EQUATION}) fulfilling the initial
conditions defined by the history set $\left\{ \widehat{\mathbf{w}}\right\}
_{t_{0}}$ are sought in the functional class of smooth 4-vector solutions of
the form $r^{\left( i\right) \mu }\equiv r^{\left( i\right) \mu }(s_{(i)})$,
with $s_{(i)}=s_{(i)}(t)$ and $t\in I_{0}\equiv \left( t_{0},\infty \right)
, $ which belong to the functional class
\begin{equation}
\left\{ \mathbf{r}(s)\right\} \equiv \left\{ \left. \left( r^{\left(
i\right) \mu },i=1,N\right) \right\vert \text{ }r^{\left( i\right) \mu
}\equiv r^{\left( i\right) \mu }(s_{(i)})\in C^{(k)}(I),\text{ with }%
s_{(i)}=s_{(i)}(t)\in C^{(k)}(I),k\geq 2,\forall t\in I_{0}\right\} .
\label{FUNCTIONAL CLASS OF SOLUTIONS}
\end{equation}%
In the following we shall assume that\emph{\ in the} \emph{setting defined
by Eq.(\ref{FUNCTIONAL CLASS OF SOLUTIONS}) with the history set (\ref%
{INITIAL HISTORY SET}),} \emph{the ODEs (\ref{N-BODY-RR-EQUATION}) admit a
unique global solution of class }$C^{(k-1)}(I_{0})$\emph{,} \emph{with} $%
k\geq 2$.

\section{$N$-body non-local Hamiltonian theory}

Based on THM.1 and its Corollary, an equivalent non-local Hamiltonian
formulation can be given for the hybrid and Lagrangian-variable approaches
stated in THM.1 and Corollary. The strategy is similar to that developed in
Paper II. Thus, first we proceed constructing the intermediate set of hybrid
variables $\mathbf{y}\equiv \left( r,p\right) \equiv \left( r^{\left(
i\right) \mu },p_{\mu }^{\left( i\right) }\text{,}i=1,N\right) $ and the
related \textit{non-local variational Hamiltonian }$H_{1}^{\left( i\right)
}=H_{1}^{\left( i\right) }(r,p,[r]),$ identified, as usual, with the
Legendre transformation of the corresponding \textit{non-local variational
Lagrangian} $L_{1}^{\left( i\right) }$. Hence, for all $i=1,N$:%
\begin{equation}
H_{1}^{\left( i\right) }=p_{\mu }^{\left( i\right) }\frac{dr^{\left(
i\right) \mu }}{ds_{\left( i\right) }}-L_{1}^{\left( i\right) },
\label{canh}
\end{equation}%
while $p_{\mu }^{\left( i\right) }$ is the $i$-th particle conjugate
momentum defined in terms of $L_{1}^{\left( i\right) }$ as%
\begin{equation}
p_{\mu }^{\left( i\right) }\equiv \frac{\partial L_{1}^{\left( i\right) }}{%
\partial \frac{dr^{\left( i\right) \mu }}{ds_{\left( i\right) }}}.
\label{canmom}
\end{equation}%
\ From THM.1 it follows that%
\begin{equation}
p_{\mu }^{\left( i\right) }=m_{o}^{\left( i\right) }cu_{\mu }^{\left(
i\right) }+\frac{q^{\left( i\right) }}{c}A_{\mu }^{(tot)\left( i\right) },
\label{pmu}
\end{equation}%
where $A_{\mu }^{(tot)\left( i\right) }$ is given by%
\begin{equation}
A_{\mu }^{(tot)\left( i\right) }(r,\left[ r\right] )=\overline{A}_{\mu
}^{(ext)\left( i\right) }+\overline{A}_{\mu }^{(self)\left( i\right) }+\sum
_{\substack{ j=1,N  \\ i\neq j}}\overline{A}_{\mu }^{(bin)\left( ij\right) },
\end{equation}%
according to the definitions given in Eqs.(\ref{Aself-i})-(\ref{Aself-ij}).
As a consequence, $H_{1}^{\left( i\right) }$ becomes simply%
\begin{equation}
H_{1}^{\left( i\right) }(r,p,[r])=\frac{1}{2m_{o}^{\left( i\right) }c}\left[
p_{\mu }^{\left( i\right) }-\frac{q^{\left( i\right) }}{c}A_{\mu
}^{(tot)\left( i\right) }\right] \left[ p^{\left( i\right) \mu }-\frac{%
q^{\left( i\right) }}{c}A^{(tot)\left( i\right) \mu }\right] .
\label{Hvariational-i}
\end{equation}%
This permits us to represent the Hamilton action functional in terms of the
hybrid state $\mathbf{y}$, yielding%
\begin{equation}
S_{H_{N}}(r,p,\left[ r\right] )=\sum_{i=1,N}S_{H_{1}^{\left( i\right) }},
\end{equation}%
where%
\begin{equation}
S_{H_{1}^{\left( i\right) }}\equiv \int_{s_{\left( i\right) 1}}^{s_{\left(
i\right) 2}}ds_{\left( i\right) }\left[ p_{\mu }^{\left( i\right) }\frac{%
dr^{\left( i\right) \mu }}{ds_{\left( i\right) }}-H_{1}^{\left( i\right) }%
\right]  \label{SH1i}
\end{equation}%
represents the $i$-th particle contribution. Of course, as an alternative,
analogous dynamical variables can be defined also in terms of the effective
Lagrangian $L_{eff}^{\left( i\right) }$ [see Eq.(\ref{Leff-i})]. This yields
the notion of \textit{effective Hamiltonian }$H_{eff}^{\left( i\right) }$ \
and of the corresponding state $\mathbf{x}\equiv \left( r,P\right) \equiv
\left( r^{\left( i\right) \mu },P_{\mu }^{\left( i\right) }\text{,}%
i=1,N\right) ,$ which will be shown below to identify a (super-abundant)
canonical state (see Corollary to THM.2). Thus, $H_{eff}^{\left( i\right) }$
- \ to be considered a non-local function of the form $H_{eff}^{\left(
i\right) }=H_{eff}^{\left( i\right) }(r,P,[r])$ - is prescribed in terms of
the Legendre transformation with respect to $L_{eff}^{\left( i\right) }$,
namely letting:%
\begin{equation}
H_{eff}^{\left( i\right) }\equiv P_{\mu }^{\left( i\right) }\frac{dr^{\left(
i\right) \mu }}{ds_{\left( i\right) }}-L_{eff}^{\left( i\right) },
\label{heff-i}
\end{equation}%
while $P_{\mu }^{\left( i\right) }$ denotes the \textit{effective canonical
momentum}%
\begin{equation}
P_{\mu }^{\left( i\right) }\equiv \frac{\partial L_{^{eff}}^{\left( i\right)
}}{\partial \frac{dr^{\left( i\right) \mu }}{ds_{\left( i\right) }}}.
\label{pleff}
\end{equation}%
From the Corollary to THM.1 it follows immediately that%
\begin{equation}
P_{\mu }^{\left( i\right) }=m_{o}^{\left( i\right) }cu_{\mu }^{\left(
i\right) }+\frac{q^{\left( i\right) }}{c}A_{\left( eff\right) \mu
}^{(tot)\left( i\right) },  \label{Peff-final}
\end{equation}%
where $A_{\left( eff\right) \mu }^{(tot)\left( i\right) }$ is given by%
\begin{equation}
A_{\left( eff\right) \mu }^{(tot)\left( i\right) }=\overline{A}_{\mu
}^{(ext)\left( i\right) }+2\overline{A}_{\mu }^{(self)\left( i\right) }+\sum
_{\substack{ j=1,N  \\ i\neq j}}\overline{A}_{\left( eff\right) \mu
}^{(bin)\left( ij\right) },  \label{A-eff-tot}
\end{equation}%
and%
\begin{equation}
\overline{A}_{\left( eff\right) \mu }^{(bin)\left( ij\right) }=2q^{\left(
j\right) }\int_{-\infty }^{+\infty }ds_{\left( j\right) }\frac{dr^{\mu
}(s_{\left( j\right) })}{ds_{\left( j\right) }}K^{\left( ij\right) },
\label{Abin-eff}
\end{equation}%
with $K^{\left( ij\right) }$ being given by Eq.(\ref{Kij}). Finally, $%
H_{eff}^{\left( i\right) }$ becomes%
\begin{equation}
H_{eff}^{\left( i\right) }=\frac{1}{2m_{o}^{\left( i\right) }c}\left[ P_{\mu
}^{\left( i\right) }-\frac{q^{\left( i\right) }}{c}A_{\left( eff\right) \mu
}^{(tot)\left( i\right) }\right] \left[ P^{\left( i\right) \mu }-\frac{%
q^{\left( i\right) }}{c}A_{\left( eff\right) }^{(tot)\left( i\right) \mu }%
\right] .  \label{heff-i2}
\end{equation}%
Therefore, by direct comparison with Eq.(\ref{Hvariational-i}) it follows
identically that
\begin{equation}
H_{eff}^{\left( i\right) }\equiv H_{1}^{\left( i\right) }.  \label{IDENTITY}
\end{equation}%
Then the following theorem, casting the Hamilton variational principle of
THM.1 in terms of the state $\mathbf{x,}$ holds.

\bigskip

\textbf{THM.2 - }$N$\textbf{-body non-local Hamiltonian variational principle%
}

\emph{Given validity of THM.1 with Corollary and the definitions (\ref{canh}%
)-(\ref{SH1i}) as well as (\ref{heff-i})-(\ref{Abin-eff}), let us assume
that the curves }$f^{\left( i\right) }(s_{\left( i\right) })\equiv \mathbf{%
\mathbf{y}}^{\left( i\right) }=(r^{\left( i\right) \mu },p_{\mu }^{\left(
i\right) })_{\left( s_{\left( i\right) }\right) }$\emph{\ belong to the
functional class }$\left\{ f\right\} $ \emph{of }$C^{2}$\emph{-functions
subject to the boundary conditions}%
\begin{equation}
\mathbf{y}^{\left( i\right) }(s_{\left( i\right) k})=\mathbf{y}_{k}^{\left(
i\right) },
\end{equation}%
\emph{for }$k=1,2,$ $s_{\left( i\right) 1},s_{\left( i\right) 2}\in
I\subseteq
\mathbb{R}
$ \emph{and with} $s_{\left( i\right) 1}<s_{\left( i\right) 2}$\emph{.}
\emph{Then the following proposition holds:}

\emph{The modified Hamilton variational principle}%
\begin{equation}
\delta S_{H_{N}}=0
\end{equation}%
\emph{subject to independent synchronous variations }$\delta f^{\left(
i\right) }(s_{\left( i\right) })\equiv \left( \delta r^{\left( i\right) \mu
}(s_{\left( i\right) }),\delta p_{\mu }^{\left( i\right) }(s_{\left(
i\right) })\right) $\emph{\ performed in the functional class indicated
above, yields, for all }$i=1,N$\emph{, the E-L equations}%
\begin{equation}
\frac{\delta S_{H_{1}^{\left( i\right) }}}{\delta p_{\mu }^{\left( i\right) }%
}=0,  \label{EL-H-1}
\end{equation}%
\begin{equation}
\frac{\delta S_{H_{1}^{\left( i\right) }}}{\delta r^{\left( i\right) \mu }}%
=0.  \label{EL-H-2}
\end{equation}%
\emph{These equations coincide identically with the }$N$\emph{-body
variational equations of motion (\ref{Eq.1RR}) and (\ref{Eq.2RR}). Hence,
the set }$\left\{ \mathbf{y},H_{N}\right\} \equiv \left\{ \mathbf{y}^{\left(
i\right) },H_{1}^{\left( i\right) },i=1,N\right\} $\emph{\ defines a
non-local Hamiltonian system.}

\emph{Proof} - The proof is analogous to that given in THM.2 of Paper II. In
particular, it can be reached, after elementary algebra, by invoking the
symmetry properties of the variational functional $S_{H_{N}}$, namely%
\begin{equation}
S_{H_{N}}(r_{A},p,\left[ r_{B}\right] )=S_{H_{N}}(r_{B},p,\left[ r_{A}\right]
),
\end{equation}%
where again $r_{A}$ and $r_{B}$ are two $N$-body arbitrary curves of the
functional class $\left\{ \mathbf{y}\right\} $. It follows that the
variational derivative in the E-L equation Eq.(\ref{EL-H-2}) becomes%
\begin{equation}
\frac{\delta S_{H_{1}^{\left( i\right) }}}{\delta r^{\left( i\right) \mu }}%
\equiv \left. \frac{\delta S_{H_{1}^{\left( i\right) }}}{\delta r^{\left(
i\right) \mu }}\right\vert _{\left[ r\right] }+\sum_{j=1,N}\left. \frac{%
\delta S_{H_{1}^{\left( j\right) }}}{\delta \left[ r^{\left( i\right) \mu }%
\right] }\right\vert _{r}=0,
\end{equation}%
where the summation is performed only on the non-local contributions. As a
consequence, the E-L equations (\ref{EL-H-1}) and (\ref{EL-H-2}) yield%
\begin{eqnarray}
&&\left. \frac{\delta S_{H_{1}^{\left( i\right) }}}{\delta p_{\mu }^{\left(
i\right) }}=m_{o}^{\left( i\right) }c\frac{dr^{\left( i\right) \mu }}{%
ds_{\left( i\right) }}-\left[ p_{\mu }^{\left( i\right) }-\frac{q^{\left(
i\right) }}{c}A_{\mu }^{(tot)\left( i\right) }\right] =0,\right. \\
&&\left. \frac{\delta S_{H_{1}^{\left( i\right) }}}{\delta r^{\left(
i\right) \mu }}=-\frac{dp_{\mu }^{\left( i\right) }}{ds_{\left( i\right) }}+%
\frac{q^{\left( i\right) }}{c}\frac{dr^{\left( i\right) \nu }(s_{i})}{%
ds_{\left( i\right) }}\left[ \frac{\partial A_{\mu }^{(tot)\left( i\right) }%
}{\partial r^{\left( i\right) \nu }}+F_{\mu \nu }^{(tot)\left( i\right) }%
\right] =0.\right.
\end{eqnarray}%
Taking into account the definitions given by Eq.(\ref{pmu}) the equivalence
with Eqs.(\ref{Eq.1RR}) and (\ref{Eq.2RR}) is immediate.

\textbf{Q.E.D.}

\bigskip

Let us now pose the problem of the construction of the corresponding $N$%
-body Hamiltonian equations in standard form, as suggested by the results of
Paper II. The non-local Hamiltonian system $\left\{ \mathbf{y},H_{N}\right\}
$ is said to admit a\emph{\ standard Hamiltonian form }$\left\{ \mathbf{x}%
,H_{eff}^{\left( 1\right) },...H_{eff}^{\left( N\right) }\right\} $\emph{\ }%
if the $N$-body equations of motion can be cast, for all $i=1,N,$ in the form%
\begin{equation}
\frac{dr^{\left( i\right) \mu }}{ds_{\left( i\right) }}=\frac{\partial
H_{eff}^{\left( i\right) }}{\partial P_{\mu }^{\left( i\right) }},
\label{HAM-1}
\end{equation}%
\begin{equation}
\frac{dP_{\mu }^{\left( i\right) }}{ds_{\left( i\right) }}=-\frac{\partial
H_{eff}^{\left( i\right) }}{\partial r^{\left( i\right) \mu }},
\label{HAM-2}
\end{equation}%
in terms of a suitably-defined \emph{effective particle Hamiltonian} $%
H_{eff}^{\left( i\right) },$ to be identified with Eq.(\ref{heff-i2}). In
particular, a $N$-body system with state $\mathbf{x}\equiv \left\{ \mathbf{x}%
^{(i)},i=1,N\right\} $ is said to be endowed with a \emph{Hamiltonian
structure }$\left\{ \mathbf{x},H_{N,eff}\right\} $ if, for all particles
belonging to the $N$-body system, the equations of motion for the $i$-th
canonical particle state $\mathbf{x}^{\left( i\right) }$ can be represented
in the PBs notation (\ref{PB}) in terms of a single Hamiltonian function $%
H_{N,eff}$, i.e., for all $i=1,N$%
\begin{equation}
\frac{d\mathbf{x}^{\left( i\right) }}{ds_{\left( i\right) }}=\left[ \mathbf{x%
}^{\left( i\right) },H_{N,eff}\right] ,  \label{ith- equation-of-motion}
\end{equation}%
with\ $H_{N,eff}$ denoting a still to be determined, appropriate \textit{%
system effective Hamiltonian}. Extending the treatment holding for the $1$%
-body problem (see Paper II), here we intend to prove that the Hamiltonian
structure $\left\{ \mathbf{x},H_{N,eff}\right\} $ holds also in the case of
EM-interacting $N$-body systems. The following proposition holds.

\bigskip

\textbf{Corollary to THM.2 - Standard Hamiltonian form and Hamiltonian
structure of the }$N$\textbf{-body equations of motion}

\emph{Given validity of THM.2 and the definitions given by Eqs.(\ref{heff-i}%
)-(\ref{heff-i2}), it follows that:}

TC2$_{1})$\emph{\ The non-local Hamiltonian system }$\left\{ \mathbf{x}%
,H_{N}\right\} $\ \emph{admits a standard Hamiltonian form defined in terms
of the set }$\left\{ \mathbf{x},H_{eff}^{\left( 1\right)
},...H_{eff}^{\left( N\right) }\right\} $\emph{, with }$H_{eff}^{\left(
i\right) }$\emph{\ the }$i$\emph{-th particle effective Hamiltonian }[given
by Eq.(\ref{heff-i2})]$;$\emph{\ furthermore} $\mathbf{x}\equiv \left\{
\mathbf{x}^{\left( i\right) },i=1,N\right\} $ \emph{is the super-abundant
canonical state, while }%
\begin{eqnarray}
\mathbf{x}^{\left( i\right) } &\equiv &\left( r^{\left( i\right) \mu
},P_{\mu }^{\left( i\right) }\right) ,  \label{CANONICAL STATE} \\
r^{\left( i\right) \mu } &=&\left( r^{\left( i\right) 0},\mathbf{r}^{\left(
i\right) }\right) ,  \label{CANONICAL POSITION} \\
P^{\left( i\right) \mu }\emph{\ } &\equiv &\left( P^{\left( i\right) 0},%
\mathbf{P}^{\left( i\right) }\right) ,  \label{CANONICAL MOMENTUM}
\end{eqnarray}%
\emph{are respectively the }$i$\emph{-th particle canonical state,
4-position and effective canonical momentum }[defined by Eq.(\ref{Peff-final}%
)]\emph{. As a consequence, Eqs.(\ref{EL-H-1}) and (\ref{EL-H-2}) can be
cast in the standard Hamiltonian form (\ref{HAM-1}) and (\ref{HAM-2}).}

TC2$_{2})$\emph{\ The equations (\ref{HAM-1}) and (\ref{HAM-2}) admit also
the equivalent representation (\ref{ith- equation-of-motion}) and hence the
set }$\left\{ \mathbf{x},H_{N,eff}\right\} $ \emph{defines a Hamiltonian
structure, with }$H_{N,eff}$ \emph{being the effective }$N$-\emph{body
Hamiltonian function}%
\begin{equation}
H_{N,eff}\equiv \sum_{i=1,N}H_{eff}^{\left( i\right) }.  \label{HN-eff-bis}
\end{equation}

TC2$_{3})$\emph{\ Introducing the system Hamiltonian }%
\begin{equation}
H_{N}\equiv \sum_{i=1,N}H^{\left( i\right) }  \label{SYSTEM HAMILTONIAN}
\end{equation}%
\emph{defined in terms of the variational }$i$\emph{-th particle variational
Hamiltonian }$H^{\left( i\right) }$ [see Eq.(\emph{\ref{Hvariational-i}})],%
\emph{\ it follows identically that }%
\begin{equation}
H_{N}=H_{N,eff}.  \label{H_N=H_Neff}
\end{equation}

\emph{Proof} - TC2$_{1})$ The proof follows from straightforward algebra.
The first equation manifestly reproduces Eq.(\ref{EL-H-1}), because of the
definition of $H_{eff}^{\left( i\right) }$ given above. Similarly, in the
second equation the partial derivative of $H_{eff}^{\left( i\right) }$
recovers the correct form of the total EM force expressed in terms of $%
A_{\left( eff\right) \mu }^{(tot)\left( i\right) }$. TC2$_{2})$\emph{\ }To
prove the existence of the Hamiltonian structure, it is sufficient to notice
that $\frac{\partial H_{eff}^{\left( i\right) }}{\partial P_{\mu }^{\left(
i\right) }}=\left[ r^{\left( i\right) \mu },H_{N,eff}\right] $ and $\frac{%
\partial H_{eff}^{\left( i\right) }}{\partial r^{\left( i\right) \mu }}=-%
\left[ P_{\mu }^{\left( i\right) },H_{N,eff}\right] $. TC2$_{3})$\ By
construction [see Eqs.(\ref{Hvariational-i}) and (\ref{heff-i2})] for all $%
i=1,N$ the effective and variational Hamiltonians coincide [see Eq.(\ref%
{IDENTITY})]$.$ This implies the validity of Eq.(\ref{H_N=H_Neff}) too,
namely $H_{N}$ identifies also the system Hamiltonian. It follows that for
all particles $i=1,N$ the canonical equations of motion (\ref{ith-
equation-of-motion}) recover the standard Hamiltonian form expressed in
terms of the PBs with respect to the the system Hamiltonian, i.e., Eqs.(\ref%
{STANDARD HAMILTONIAN FORM}), so that $\left\{ \mathbf{x,}H_{N}\right\} $
identifies the Hamiltonian structure of the EM-interacting $N$-body system.

\textbf{Q.E.D.}

\bigskip

\section{General implications of the non-local $N$-body theory}

Let us now comment on the general implications of the previous theorems.

\emph{Remark \#1 - Difference form of the Hamilton equations of motion. }The
canonical equations (\ref{STANDARD HAMILTONIAN FORM}) imply the following
difference equations, i.e., the infinitesimal canonical transformation
generated by $H_{N}$:%
\begin{equation}
d\mathbf{x}^{\left( i\right) }=ds_{\left( i\right) }\left[ \mathbf{x}%
^{\left( i\right) },H_{N}\right] .  \label{DIFFERENCE EQS.}
\end{equation}

\emph{Remark \#2 - Coordinate-time representation of the Hamilton equations
of motion. }Introducing the coordinate-time parametrization (\ref%
{COORDINATE-TIME PARAMETRIZATION}), Eqs.(\ref{DIFFERENCE EQS.}) become%
\begin{equation}
d\mathbf{x}^{\left( i\right) }=\frac{cdt}{\gamma _{\left( i\right) }}\left[
\mathbf{x}^{\left( i\right) },H_{N}\right] ,
\label{COORDINATE-TIME DIFFERENCE EQS.}
\end{equation}%
with $\gamma _{(i)}$ denoting again the relativistic factor (\ref{GAMMA_i}),
while $\left[ \cdot ,\cdot \right] $ are the local PBs evaluated with
respect to the super-abundant canonical state $\mathbf{x}$. These yield
explicitly%
\begin{equation}
dr^{\left( i\right) \mu }=\frac{dt}{\gamma _{\left( i\right) }}\frac{1}{%
m_{o}^{\left( i\right) }}\left( P_{\mu }^{\left( i\right) }-\frac{q^{\left(
i\right) }}{c}A_{\left( eff\right) \mu }^{(tot)\left( i\right) }\right)
\equiv \frac{dt}{\gamma _{\left( i\right) }}cu_{\mu }^{(i)}\equiv dtv_{\mu
}^{(i)},  \label{DIFFERENCE EQ-1}
\end{equation}%
\begin{equation}
dP_{\mu }^{(i)}=\frac{dt}{\gamma _{\left( i\right) }}\frac{q^{\left(
i\right) }}{m_{o}^{\left( i\right) }c}\frac{\partial A_{\left( eff\right)
\nu }^{(tot)\left( i\right) }}{\partial r^{\left( i\right) \mu }}\left(
P^{\left( i\right) \nu }-\frac{q^{\left( i\right) }}{c}A_{\left( eff\right)
}^{(tot)\left( i\right) \nu }\right) \equiv \frac{cdt}{\gamma _{\left(
i\right) }}\frac{q^{\left( i\right) }}{c}\frac{\partial A_{\left( eff\right)
\nu }^{(tot)\left( i\right) }}{\partial r^{\left( i\right) \mu }}u^{(i)\nu
}\equiv dt\frac{q^{\left( i\right) }}{c}\frac{\partial A_{\left( eff\right)
\nu }^{(tot)\left( i\right) }}{\partial r^{\left( i\right) \mu }}v^{(i)\nu }.
\label{DIFFERENCE EQ-2}
\end{equation}%
\emph{Remark \#3 - Well-posedness of the N-body equations of motion. }All
the equations of motion indicated above [see THMs.1 and 2 and their
Corollaries] are equivalent to each other and are manifestly
Lorentz-covariant (see also related discussion in Paper II). Then, a
well-posed problem for the Hamiltonian equations in standard form (\ref{ith-
equation-of-motion}) can be obtained in analogy to the problem defined by
Eqs.(\ref{N-BODY-RR-EQUATION}),(\ref{INITIAL HISTORY SET}). This is
achieved, first, by prescribing the appropriate initial history set\emph{\ }$%
\left\{ \widehat{\mathbf{x}}\right\} _{t_{0}}\subset \Gamma _{N}$\emph{. }%
For an arbitrary coordinate initial time $t_{0}\in I\equiv
\mathbb{R}
,$ this is defined as the ensemble of initial states
\begin{equation}
\left\{ \widehat{\mathbf{x}}\right\} _{t_{0}}\equiv \left\{ \widehat{\mathbf{%
x}}(t)\equiv \left( \left( \widehat{r}^{(i)}(t),\widehat{p}^{(i)}(t)\right)
\in C^{(k-1)}(I),i=1,N\right) ,\forall t\in \left[ t_{0}-t_{ret}^{\max
}(t_{0}),t_{0}\right] ,k\geq 2\right\}  \label{CANONICAL HISTORY SET}
\end{equation}%
(\emph{canonical history set}), where $t_{ret}^{\max }(t_{0})$ is the
maximum delay-time at $t_{0}$ [see Eq.(\ref{MAX-DELAY TIME})]. Furthermore,
in analogy with Eq.(\ref{FUNCTIONAL CLASS OF SOLUTIONS}), solutions of Eqs.(%
\ref{ith- equation-of-motion}) fulfilling the initial conditions defined by
the history set $\left\{ \widehat{\mathbf{x}}\right\} _{t_{0}}$ are sought
in the functional class (\ref{FUNCTIONAL CLASS}) by identifying
\begin{equation}
\emph{\ }f^{\left( i\right) }(s_{\left( i\right) })\equiv \left[ r^{\left(
i\right) \mu }(s_{\left( i\right) }),p_{\mu }^{\left( i\right) }(s_{\left(
i\right) })\right] .  \label{CANONICAL FUNCTIONAL SETTING}
\end{equation}%
In the following we shall assume that\emph{\ in the} \emph{setting defined
by Eq.(\ref{CANONICAL FUNCTIONAL SETTING}) with the canonical history set (%
\ref{CANONICAL HISTORY SET}),} \emph{the ODEs (\ref{ith- equation-of-motion}%
) admit a unique global solution of class }$C^{(k-1)}(I_{0})$ \emph{with} $%
k\geq 2$.

\emph{Remark \#4 - Extremant and extremal curves.} In all cases indicated
above the solutions of $N$-body E-L equations of motion (\textit{extremal
curves}) and of the Lagrangian and Hamiltonian equations in standard form
given by the Corollaries to THMs.1 and 2 (\textit{extremant curves}),
satisfy identically, for all $i=1,N$, the kinematic constraints
\begin{equation}
u_{\mu }^{\left( i\right) }u^{\left( i\right) \mu }=1
\label{VELOCITY KINEMATIC CONSTRAINT}
\end{equation}%
(\textit{velocity constraints}) and
\begin{equation}
ds_{\left( i\right) }^{2}=\eta _{\mu \nu }dr^{\left( i\right) \mu
}dr^{\left( i\right) \nu }  \label{LINE-ELEMENT KINEMATIC--CONSTRAINT}
\end{equation}%
(\textit{line-element constraints}). The first constraint implies that the
time-components of the 4-velocity depend on the corresponding space
components, while the second constraint requires that the particles' proper
times are uniquely related to the corresponding coordinate times.\ In
particular, we shall denote as \textit{extremant canonical curves}%
\begin{equation}
\mathbf{x}(s_{(1)},...s_{(N)})\equiv \left\{ \mathbf{x}^{(1)}(s_{(1)}),...%
\mathbf{x}^{(N)}(s_{(N)})\right\} ,  \label{EXTREMANT CANONICAL-CURVES}
\end{equation}%
with $\mathbf{x}^{(i)}=\mathbf{x}^{(i)}(s_{(i)})$ for all $i=1,N,$ arbitrary
particular solutions of the canonical equations (\ref{CANONICAL EQUATIONS}).

\emph{Remark \#5 - Unconstrained varied functions.} By assumption, both the
varied functions\emph{\ }$f^{\left( i\right) }=\left[ r^{\left( i\right) \mu
},u_{\mu }^{\left( i\right) }\right] _{(s_{\left( i\right) })},$ $\mathbf{%
\mathbf{y}}^{\left( i\right) }=(r^{\left( i\right) \mu },p_{\mu }^{\left(
i\right) })_{\left( s_{\left( i\right) }\right) }$ and $\mathbf{\mathbf{x}}%
^{\left( i\right) }=(r^{\left( i\right) \mu },P_{\mu }^{\left( i\right)
})_{\left( s_{\left( i\right) }\right) }$ entering respectively THMs.1 and 2
as well as the Corollary of THM.2 are \emph{unconstrained}, namely they are
solely subject to the requirement that end points and boundary values are
kept fixed (and therefore do \textit{not} fulfill the previous kinematic
constraints). This implies, in particular, that all of the ($8$) components
of $f^{\left( i\right) },$ $\mathbf{\mathbf{y}}^{\left( i\right) }$ and $%
\mathbf{\mathbf{x}}^{\left( i\right) }$ must be considered independent. On
the other hand, both the extremal and extremant curves satisfy all of the
required kinematic constraints, so that only (6) of them are actually
independent for each particle (see also discussion in Paper I).

\emph{Remark \#6 - Non-local Hamiltonian structure and unconstrained
canonical state. }Thanks to proposition TC2$_{3}$ of the Corollary to THM.2
the Hamiltonian structure $\left\{ \mathbf{x},H_{N,eff}\right\} $ coincides
with $\left\{ \mathbf{x},H_{N}\right\} ,$ $H_{N}$ denoting the non-local
system Hamiltonian defined by Eq.(\ref{SYSTEM HAMILTONIAN}).\emph{\ }We
remark, however, that in the PBs given by Eq.(\ref{STANDARD HAMILTONIAN FORM}%
) the partial derivatives must be evaluated with respect to the
unconstrained states $\mathbf{x}^{\left( i\right) }$ and not $\mathbf{y}%
^{\left( i\right) }$ indicated above. This means that $H_{N}$ must be
considered a function of $\mathbf{x}$. It is immediate to prove that the
same Hamiltonian structure $\left\{ \mathbf{x},H_{N}\right\} $ holds
provided the super-abundant canonical state $\mathbf{x}\equiv \left( \mathbf{%
x}^{\left( i\right) },i=1,N\right) $ is considered unconstrained. In fact,
as shown by the Corollary to THM.2, in such a case the canonical equations
in standard form (\ref{HAM-1}) and (\ref{HAM-2}) admit a PB-representation
of the form (\ref{ith- equation-of-motion}). For this purpose let us make
use of the coordinate-time parametrization $\mathbf{x}\equiv \widehat{%
\mathbf{x}}(t),$ denoting $\widehat{\mathbf{x}}(t)=\mathbf{x}_{o}+d\mathbf{x}
$, with $\mathbf{x}_{o}\equiv \widehat{\mathbf{x}}(t_{o})$ and $d\mathbf{x}%
\equiv \left( d\mathbf{x}^{\left( 1\right) },...d\mathbf{x}^{\left( N\right)
}\right) .$ Furthermore let us require that the initial history set\emph{\ }$%
\left\{ \widehat{\mathbf{x}}\right\} _{t_{0}}$ is prescribed. Then, a
necessary and sufficient condition for the equations of motion to admit the
standard Hamiltonian form (\ref{STANDARD HAMILTONIAN FORM}) is that the
fundamental local PBs for the state $\mathbf{x}\equiv \widehat{\mathbf{x}}%
(t) $, defined with respect to the same state $\mathbf{x}_{o}\equiv \widehat{%
\mathbf{x}}(t_{o})$,\ namely%
\begin{eqnarray}
\left[ r^{\left( i\right) \mu },r^{\left( j\right) \nu }\right] _{(\mathbf{x}%
_{o})} &=&0,  \notag \\
\left[ P_{\mu }^{\left( i\right) },P_{\nu }^{\left( j\right) }\right] _{(%
\mathbf{x}_{o})} &=&0,  \notag \\
\left[ r^{\left( i\right) \mu },P_{\nu }^{\left( j\right) }\right] _{(%
\mathbf{x}_{o})} &=&\delta ^{ij}\delta _{\nu }^{\mu },  \label{Fund-PB}
\end{eqnarray}%
are identically satisfied for all $i,j=1,N$ and $\mu ,\nu =0,3$. This is
realized only when\emph{\ }the super-abundant variables $\mathbf{x}\equiv
\left( \mathbf{x}^{\left( i\right) },i=1,N\right) $ \emph{are considered
independent}$.$

\emph{Remark \#7 - The canonical flow is not a dynamical system. }A final
issue concerns the properties of the flow generated by the canonical problem
(\ref{ith- equation-of-motion}) and (\ref{CANONICAL HISTORY SET})-(\ref%
{CANONICAL FUNCTIONAL SETTING}) (\emph{canonical flow}). In the $N$-body
phase-space $\Gamma _{N}$ this is an ensemble $C^{(k-1)}-$homeomorphism
(with $k\geq 2$) of the type%
\begin{equation}
\left\{ \widehat{\mathbf{x}}\right\} _{t_{0}}\leftrightarrow \widehat{%
\mathbf{x}}(t),  \label{FLOW}
\end{equation}%
which maps an arbitrary history set $\left\{ \widehat{\mathbf{x}}\right\}
_{t_{0}}\subset \Gamma _{N}$ onto a state $\widehat{\mathbf{x}}(t)$ crossed
at a later coordinate time $t$ (i.e., at $t>t_{0}$). \emph{This map does not
generally define a dynamical system.} In fact, unless there is a subset on
non-vanishing measure of $\Gamma _{N}$ in which $\left\{ \widehat{\mathbf{x}}%
\right\} _{t_{0}}$ reduces to the initial \emph{instant set}%
\begin{equation}
\left\{ \widehat{\mathbf{x}}\right\} _{t_{0}}\equiv \left\{ \widehat{\mathbf{%
x}}\mathbf{(}t_{o})=\mathbf{x}_{0}\right\} ,  \label{INSTANT SET}
\end{equation}%
the flow (\ref{FLOW}) is not a bijection in $\Gamma _{N}$. To prove the
statement it is sufficient to notice that - in the case of a non-local
Hamiltonian structure $\left\{ \mathbf{x},H_{N,eff}\right\} $ - if the
history set is left unspecified and only the initial state $\widehat{\mathbf{%
x}}(t_{o})$ is prescribed, the image of the initial state is obviously
generally non-unique [and hence it may not coincide with $\widehat{\mathbf{x}%
}(t)$]. In fact, while the same initial state $\widehat{\mathbf{x}}(t_{o})$
may be produced by different history sets, for example $\left\{ \widehat{%
\mathbf{x}}\right\} _{t_{0}}$ and $\left\{ \widehat{\mathbf{x}}^{\prime
}\right\} _{t_{0}},$ the same history sets will generally give rise to
different images $\widehat{\mathbf{x}}(t)$ and $\widehat{\mathbf{x}}^{\prime
}(t).$

We emphasize that for $N$-body systems subject to EM interactions the
instant set (\ref{INSTANT SET}) can be realized only for special initial
conditions, i.e., for example, if for all $t\leq t_{0}$ all the particles of
the system are in inertial motion with respect to an inertial Lorentz frame.
Since, unlike the external EM field, binary and self EM interactions cannot
be \textquotedblleft turned off\textquotedblright , it follows that the set
of initial conditions (\ref{INSTANT SET}) has necessarily null measure in $%
\Gamma _{N}$.

\bigskip

\bigskip

\section{The $N$-body Hamiltonian asymptotic approximation}

In this section we want to develop asymptotic approximations for the
equations of motion of EM-interacting $N$-body systems. This involves
different asymptotic conditions to be imposed on both the self and binary
interactions. In particular:

1) For the RR self-interaction of each particle $i$ with itself this is
provided by the \emph{short delay-time ordering}, namely the requirement
that the dimensionless parameters $\epsilon _{\left( i\right) }\equiv \frac{%
(s_{\left( i\right) }-s_{\left( i\right) }^{\prime })}{s_{\left( i\right) }}%
, $ for $i=1,N$ are all infinitesimal of the same order $\epsilon $, i.e. $%
\epsilon \sim \epsilon _{\left( i\right) }\ll 1$, $s_{\left( i\right)
}-s_{\left( i\right) }^{\prime }$ denoting the $i$-th proper-time difference
between observation ($s$) and emission ($s^{\prime }$) of self-radiation.

2) For the binary EM interactions the Minkowski distance $\left\vert
\widetilde{R}^{\left( ij\right) \alpha }\right\vert $ between two arbitrary
particles of the system is much larger than their radii, in the sense that
for all $i,j=1,N$, with $i\neq j$, the \emph{large-distance ordering }$0<%
\frac{\sigma _{\left( i\right) }}{\left\vert \widetilde{R}^{\left( ij\right)
\alpha }\right\vert }\sim \frac{\sigma _{\left( j\right) }}{\left\vert
\widetilde{R}^{\left( ij\right) \alpha }\right\vert }\lesssim \epsilon $
holds.

The fundamental issue arises whether an approximation can be found for the $%
N $-body problem which:

1) is consistent with the orderings 1) and 2);

2) recovers the variational, Lagrangian and Hamiltonian character of the
exact theory (see THMs.1 and 2);

3) preserves both the standard Lagrangian and Hamiltonian forms of the
equations of motion;

4) retains finite delay-time effects characteristics of both the RR and
binary EM interactions, consistent with the prerequisites \#1-\#5 indicated
above.

In this regard, a fundamental result is the discovery pointed out in Paper
II of an asymptotic Hamiltonian approximation of this type for single
extended particles subject to the EM self-interaction. This refers to the
retarded-time Taylor expansion of the Faraday tensor contribution carried by
the RR self-force. More precisely, in the case of a single particle, this is
obtained by Taylor-expanding the RR self-force%
\begin{equation}
G_{\mu }^{\left( i\right) }\equiv \frac{q^{\left( i\right) }}{c}\overline{F}%
_{\mu k}^{\left( self\right) \left( i\right) }\left( r^{\left( i\right)
}\left( s_{\left( i\right) }\right) ,r^{\left( i\right) }\left( s_{\left(
i\right) }^{\prime }\right) \right) \frac{dr^{\left( i\right) k}(s_{\left(
i\right) })}{ds_{\left( i\right) }}  \label{Gmu-i}
\end{equation}%
for $i=1$ (see Eq.(\ref{COVARIANT FORM})) \textit{in the neighborhood of the
retarded proper-time }$s_{\left( i\right) }^{\prime }$. Here we claim that
an analogous conclusion can be drawn also for the corresponding $N$-body
problem, by introducing the same expansion to all charged particles and
invoking the large-distance ordering for the binary interaction. For this
purpose, we shall assume that the external force acting on each charged
particle is slowly varying in the sense that, denoting $r^{\prime }\equiv
r^{\left( i\right) \mu }\left( s_{\left( i\right) }^{\prime }\right) $ and $%
r\equiv r^{\left( i\right) \mu }\left( s_{\left( i\right) }\right) $,%
\begin{eqnarray}
\overline{F}_{\mu \nu }^{(ext)}\left( r^{\prime }\right) -\overline{F}_{\mu
\nu }^{(ext)}\left( r\right) &\sim &O\left( \epsilon \right) ,
\label{smooth1} \\
\left( \overline{F}_{\mu \nu }^{(ext)}\left( r^{\prime }\right) -\overline{F}%
_{\mu \nu }^{(ext)}\left( r\right) \right) _{,h} &\sim &O\left( \epsilon
\right) , \\
\left( \overline{F}_{\mu \nu }^{(ext)}\left( r^{\prime }\right) -\overline{F}%
_{\mu \nu }^{(ext)}\left( r\right) \right) _{,hk} &\sim &O\left( \epsilon
\right) .  \label{smooth3}
\end{eqnarray}

Then, the following proposition holds.

\bigskip

\textbf{THM.3 - }$N$\textbf{-body asymptotic Hamiltonian approximation.}

\emph{Given validity of THM.2 and the short delay-time and large-distance
asymptotic orderings as well as the smoothness assumptions (\ref{smooth1})-(%
\ref{smooth3}) for the external EM field, neglecting corrections of order }$%
\epsilon ^{n},$ \emph{with} $n\geq 1$ \emph{(first-order approximation), the
following results hold:}

\emph{T3}$_{1}$\emph{) The vector fields (\ref{Gmu-i}) describing the RR
self-force are approximated in a neighborhood of }$s_{\left( i\right)
}^{\prime }$ \emph{as}%
\begin{equation}
g_{\mu }^{\left( i\right) }\left( r^{\left( i\right) }\left( s_{\left(
i\right) }^{\prime }\right) \right) =\left\{ -m_{oEM}^{\left( i\right) }c%
\frac{d}{ds_{\left( i\right) }^{\prime }}u_{\mu }^{\left( i\right) }\left(
s_{\left( i\right) }^{\prime }\right) +g_{\mu }^{\left( i\right) \prime
}\left( r^{\left( i\right) }\left( s_{\left( i\right) }^{\prime }\right)
\right) \right\} ,  \label{asymp}
\end{equation}%
\emph{to be referred to as retarded-time Hamiltonian approximation for the
self-force, in which the first term on the r.h.s. identifies a retarded
mass-correction term, }$m_{oEM}^{\left( i\right) }\equiv \frac{q^{\left(
i\right) 2}}{c^{2}\sigma _{\left( i\right) }}$ \emph{denoting the
leading-order EM mass. Finally, }$g_{\mu }^{\left( i\right) \prime }$\emph{\
are the 4-vectors}%
\begin{equation}
g_{\mu }^{\left( i\right) \prime }\left( r^{\left( i\right) }\left(
s_{\left( i\right) }^{\prime }\right) \right) =-\frac{1}{3}\frac{q^{\left(
i\right) 2}}{c}\left[ \frac{d^{2}}{ds_{\left( i\right) }^{\prime 2}}u_{\mu
}^{\left( i\right) }\left( s_{\left( i\right) }^{\prime }\right) -u_{\mu
}^{\left( i\right) }(s_{\left( i\right) }^{\prime })u^{\left( i\right)
k}(s_{\left( i\right) }^{\prime })\frac{d^{2}}{ds_{\left( i\right) }^{\prime
2}}u_{k}^{\left( i\right) }\left( s_{\left( i\right) }^{\prime }\right) %
\right] .
\end{equation}

\emph{T3}$_{2}$\emph{) The tensor fields }$\overline{F}_{\mu \nu }^{\left(
bin\right) \left( ij\right) }$ \emph{for all }$i,j=1,N$\emph{, with }$i\neq
j $\emph{, appearing in the binary EM interaction (see Eq.(\ref{Fbin-ij}))
are approximated by the leading-order (point-particle) terms:}%
\begin{equation}
\overline{F}_{\mu \nu }^{\left( bin\right) \left( ij\right) }\cong \overline{%
F}_{\mu \nu }^{\left( bin\right) \left( ij\right) }\left( r^{\left( i\right)
},\left[ r^{\left( j\right) }\right] ,\sigma _{\left( i\right) }=0\text{%
\emph{,}}\sigma _{\left( j\right) }=0\right) .  \label{asymp-2}
\end{equation}

\emph{T3}$_{3}$\emph{) The corresponding asymptotic }$N$\emph{-body
equations of motion obtained replacing }$G_{\mu }^{\left( i\right) }$\emph{\
and }$\overline{F}_{\mu \nu }^{\left( bin\right) \left( ij\right) }$\emph{\
with the asymptotic approximations (\ref{asymp}) and (\ref{asymp-2}) are
variational, Lagrangian and admit a standard Lagrangian form. Denoting with }%
$r_{0}^{\left( i\right) \prime }\equiv r_{0}\left( s_{\left( i\right)
}^{\prime }\right) $ \emph{the extremal }$i$\emph{-th particle world-line at
the retarded proper time }$s_{\left( i\right) }^{\prime }$\emph{, the }$i$%
\emph{-th particle} \emph{asymptotic variational Lagrangian functions become:%
}

\begin{equation}
L_{1,asym}^{\left( i\right) }(r,u,\left[ r\right] )=L_{M}^{\left( i\right)
}(r,u)+L_{C}^{(ext)\left( i\right) }(r)+L_{C,asym}^{(self)\left( i\right)
}(r^{\left( i\right) },r_{0}^{\left( i\right) \prime
})+L_{C,asym}^{(bin)\left( i\right) }(r,\left[ r\right] ).  \label{L1-asym}
\end{equation}%
\emph{Here }$L_{M}^{\left( i\right) }$\emph{\ and }$L_{C}^{(ext)\left(
i\right) }$\emph{\ remain unchanged (see Eqs.(\ref{Lmass-ij}) and (\ref%
{Lext-ij})), while the non-local terms }$L_{C,asym}^{(self)\left( i\right)
}(r,\left[ r\right] )$\emph{\ and }$L_{C,asym}^{(bin)\left( i\right) }(r,%
\left[ r\right] )$\emph{\ are respectively}%
\begin{eqnarray}
L_{C,asym}^{(self)\left( i\right) }(r^{\left( i\right) },r_{0}^{\left(
i\right) \prime }) &=&g_{\mu }^{\left( i\right) }\left( r_{0}^{\left(
i\right) \prime }\right) r^{\left( i\right) \mu }, \\
L_{C,asym}^{(bin)\left( i\right) }(r,\left[ r\right] ) &=&\frac{q^{\left(
i\right) }}{c}\frac{dr^{\left( i\right) \mu }}{ds_{\left( i\right) }}\sum
_{\substack{ j=1,N  \\ i\neq j}}\overline{A}_{\mu }^{(bin)\left( ij\right)
}\left( \sigma _{\left( j\right) }=0\right) ,
\end{eqnarray}%
\emph{where, from Eq.(\ref{Aself-ij}) one obtains}%
\begin{equation}
\overline{A}_{\mu }^{(bin)\left( ij\right) }\left( \sigma _{\left( j\right)
}=0\right) \equiv 2q^{\left( j\right) }\int_{-\infty }^{+\infty }ds_{\left(
j\right) }\frac{dr^{\mu }(s_{\left( j\right) })}{ds_{\left( j\right) }}%
\delta (\widetilde{R}^{\left( ij\right) \alpha }\widetilde{R}_{\alpha
}^{\left( ij\right) }).
\end{equation}%
\emph{Similarly, the effective particle Lagrangians are, for }$i=1,N$\emph{:}%
\begin{equation}
L_{eff,asym}^{\left( i\right) }(r,u,\left[ r\right] )\equiv L_{M}^{\left(
i\right) }(r,u)+L_{C}^{(ext)\left( i\right) }(r)+L_{C,asym}^{(self)\left(
i\right) }(r^{\left( i\right) },r_{0}^{\left( i\right) \prime
})+2L_{C,asym}^{(bin)\left( i\right) }(r,\left[ r\right] ).
\label{Leff-asym}
\end{equation}

\emph{T3}$_{4}$\emph{) The }$N$\emph{-body equations obtained imposing the
asymptotic approximations given by Eqs.(\ref{asymp}) and (\ref{asymp-2}) are
also Hamiltonian. The asymptotic variational and effective Hamiltonian
functions are given respectively by}%
\begin{eqnarray}
H_{1,asym}^{\left( i\right) } &=&p_{\mu }^{\left( i\right) }\frac{dr^{\left(
i\right) \mu }}{ds_{\left( i\right) }}-L_{1,asym}^{\left( i\right) }, \\
H_{eff,asym}^{\left( i\right) } &\equiv &P_{\mu }^{\left( i\right) }\frac{%
dr^{\left( i\right) \mu }}{ds_{\left( i\right) }}-L_{eff,asym,}^{\left(
i\right) }  \label{Heff-asym}
\end{eqnarray}%
\emph{with }$L_{1,asym}^{\left( i\right) }$\emph{\ and }$L_{eff,asym,}^{%
\left( i\right) }$\emph{\ defined by Eqs.(\ref{L1-asym}) and (\ref{Leff-asym}%
), while now}%
\begin{eqnarray}
p_{\mu }^{\left( i\right) } &\equiv &\frac{\partial L_{1,asym}^{\left(
i\right) }}{\partial \frac{dr_{\mu }^{\left( i\right) }(s_{\left( i\right) })%
}{ds_{\left( i\right) }}}, \\
P_{\mu }^{\left( i\right) } &\equiv &\frac{\partial L_{^{eff,asym}}^{\left(
i\right) }}{\partial \frac{dr^{\left( i\right) \mu }}{ds_{\left( i\right) }}}%
.
\end{eqnarray}

\emph{Proof} - T3$_{1}$) The proof is analogous to that given in THM.5 of
Paper II. T3$_{2}$) To prove the validity of Eq.(\ref{asymp-2}), let us
recall the definition of $\overline{F}_{\mu \nu }^{\left( bin\right) \left(
ij\right) }$ given by Eq.(\ref{Fbin-ij}). Then, for each particle, imposing
the large-distance ordering and neglecting corrections of order $\epsilon
^{n},$ with $n\geq 1$, the leading-order contribution is given by Eq.(\ref%
{asymp-2}), which depends on a single delay-time determined by the positive
root of the equation $\widetilde{R}^{\left( ij\right) \alpha }\widetilde{R}%
_{\alpha }^{\left( ij\right) }=0.$ T3$_{3}$) The proof follows by first
noting that the function $L_{C,asym}^{(self)\left( i\right) }$ contributes
to the $i$-th particle E-L equations only in terms of the local dependence
in terms of $r^{\left( i\right) }$. Second, in the large-distance ordering,
the asymptotic approximation for the $N$-body Lagrangian carrying the binary
interactions yields a symmetric functional, as the exact one. Therefore, the
$N$-body asymptotic equations are necessarily variational and Lagrangian.
Straightforward algebra shows that the E-L equations determined with respect
to the asymptotic variational Lagrangian (\ref{L1-asym}) coincide with the
asymptotic approximations proved by propositions T3$_{1}$) and T3$_{2}$). In
a similar way it is immediate to prove the validity of Eq.(\ref{Leff-asym}),
which shows that the same asymptotic equations admit a standard Lagrangian
form. T3$_{4}$) Finally, the equivalent $N$-body variational and standard
Hamiltonian formulations follow by performing Legendre transformations on
the corresponding asymptotic variational and effective Lagrangian functions.
It follows that the asymptotic $N$-body equations of motion can also be
represented in the standard Hamiltonian form in terms of $%
H_{eff,asym}^{\left( i\right) }$.

\textbf{Q.E.D.}

\bigskip

It is worth pointing out the unique features of THM.3. These are related, in
particular, to the asymptotic expansion performed on the RR self-force
alone. In most of the previous literature, the short delay-time expansion is
performed with respect to the particle present proper-time. This leads
unavoidably to local asymptotic equations (analogous to the LAD and LL
equations) which are intrinsically non-variational and therefore
non-Lagrangian and non-Hamiltonian. In contrast, the short delay-time
expansion adopted here (as in Paper II) approximates the non-local RR vector
field in a manner that meets the goals indicated at the beginning of the
section. The remarkable consequence is that the asymptotic $N$-body
equations of motion retain the representation in standard Hamiltonian form
characteristic of the corresponding exact equations. Finally, we stress that
in all cases both for the exact and asymptotic formulations, the variational
and effective Lagrangian and Hamiltonian functions are always non-local
functions of the particle states. The non-locality is intrinsic and arises
even in the 1-body systems, being due to the functional form of the EM
4-potential generated by each extended particle.

\bigskip

\section{On the validity of Dirac generator formalism}

A seminal approach in relativistic dynamics is the Dirac generator formalism
developed originally by Dirac (Dirac, 1949 \cite{Dirac}) to describe the
dynamics of interacting $N$-body systems in the Minkowski space-time.
Dirac's primary goal is actually to determine the underlying \emph{dynamical
system}, exclusively based on DGF. In his words \textquotedblleft \textit{In
setting up such a new dynamical system one is faced at the outset by the two
requirements of special relativity and of Hamiltonian equations of
motion\textquotedblright }. It thus \textquotedblleft ...\textit{becomes a
matter of great importance to set up\ }(in this way) \textit{new dynamical
systems and see if they will better describe the atomic world}%
\textquotedblright\ (quoted from Ref.\cite{Dirac}).

DGF is couched on the Lie algebra of Poincar\`{e} generators for\ classical $%
N$-body systems. The basic hypothesis behind Dirac approach is that these
systems must have a Hamiltonian structure $\left\{ \mathbf{z},K_{N}\right\} $
of some sort, with $\mathbf{z}=\left\{ \mathbf{z}^{\left( 1\right) },...%
\mathbf{z}^{\left( N\right) }\right\} $ and $K_{N}$ being a suitable
canonical state and a Hamiltonian function of the system. In particular, the
canonical states $\mathbf{z}^{\left( i\right) }$ of all particles $i=1,N$
must satisfy, by assumption, covariant Hamilton equations of motion of the
form%
\begin{equation}
\frac{d\mathbf{z}^{\left( i\right) }}{ds_{\left( i\right) }}=\left[ \mathbf{z%
}^{\left( i\right) },K_{N}\right] ,  \label{CANONICAL EQUATIONS}
\end{equation}%
with $s_{\left( i\right) }$ denoting the $i$-th particle proper time.
However, it must\ be stressed that certain aspects of Dirac theory remain
\textquotedblleft a priori\textquotedblright\ undetermined. This concerns
the functional settings both of the canonical state $\mathbf{z}$ and of the
Hamiltonian function $K_{N}.$ In particular, the system state $\mathbf{z}%
=\left\{ \mathbf{z}^{\left( 1\right) },...\mathbf{z}^{\left( N\right)
}\right\} $ remains in principle unspecified, so that it might be identified
either with a set of \emph{super-abundant }or\emph{\ essential} canonical
variables. Thus, for example, in the two cases the $i$-the particle state $%
\mathbf{z}^{\left( i\right) }$ might be prescribed respectively either as:
a) the ensemble of two $4$-vectors $\mathbf{z}^{\left( i\right) }=\left(
r^{(i)\mu },\pi ^{(i)\mu }\right) ,$ with $r^{(i)\mu }=\left( r^{(i)0},%
\mathbf{r}^{(i)}\right) $ and $\pi ^{(i)\mu }=\left( \pi ^{(i)0},\mathbf{\pi
}^{(i)}\right) $ being respectively the particle $4$-position and its
conjugate $4$-momentum; b) the ensemble of the corresponding two $3$-vectors
obtained taking only the space parts of the same $4$-vectors $r^{(i)\mu }$
and $\pi ^{(i)\mu },$ namely in terms of $\mathbf{z}^{\prime \left( i\right)
}=\left( \mathbf{r}^{(i)},\mathbf{\pi }^{(i)}\right) .$ Thus, depending on
the possible prescription, the very definitions of the PBs entering Eqs.(\ref%
{CANONICAL EQUATIONS}) and DGF change.

Furthermore,\ it remains \textquotedblleft a priori\textquotedblright\
unspecified whether $K_{N}$ is actually intended as a local or a non-local
function of the canonical state $\mathbf{z}$. In Ref.\cite{Dirac}, however,
certain restrictions on the nature of the set $\left\{ \mathbf{z}%
,K_{N}\right\} $ are actually implied. These will be discussed below,
leaving aside for the moment further discussions on this important issue.

Provided the Hamiltonian structure $\left\{ \mathbf{z},K_{N}\right\} $
exists, any set of smooth dynamical variables $\eta ,$ $\xi $ and $\zeta $
depending locally on the canonical state $\mathbf{z}$ necessarily fulfills
the following laws%
\begin{eqnarray}
&&\left. \left[ \eta ,\xi \right] =-\left[ \xi ,\eta \right] ,\right.  \notag
\\
&&\left. \left[ \eta ,\xi +\zeta \right] =\left[ \eta ,\xi \right] +\left[
\eta ,\zeta \right] ,\right.  \notag \\
&&\left. \left[ \xi ,\eta \zeta \right] =\left[ \xi ,\eta \right] \zeta
+\eta \left[ \xi ,\zeta \right] ,\right.  \notag \\
&&\left. \left[ \left[ \xi ,\eta \right] ,\zeta \right] +\left[ \left[ \eta
,\zeta \right] ,\xi \right] +\left[ \left[ \zeta ,\xi \right] ,\eta \right]
=0.\right.
\end{eqnarray}%
DGF relies on the Lie transformation formalism and is based on the
representation of the Lorentz transformation group in terms of the
corresponding generator algebra. This is defined as the set of
phase-functions (Poincar\`{e} algebra generators) $\left\{ F\right\} $ given
by%
\begin{equation}
F=-\widehat{p}^{\mu }a_{\mu }+\frac{1}{2}\widehat{M}^{\mu \nu }b_{\mu \nu },
\label{Poincare-generator}
\end{equation}%
with $a_{\mu }$, $b_{\mu \nu }$ being suitable real constant infinitesimals
and $\widehat{p}^{\mu },$ $\widehat{M}^{\mu \nu }=-\widehat{M}^{\nu \mu }$
appropriate local phase-functions obeying the PBs (\emph{Lorentz conditions})%
\begin{eqnarray}
\left[ \widehat{p}_{\mu },\widehat{p}_{\nu }\right] &=&0,  \notag \\
\left[ \widehat{M}_{\mu \nu },\widehat{p}_{\alpha }\right] &=&-\eta _{\mu
\alpha }\widehat{p}_{\nu }+\eta _{\nu \alpha }\widehat{p}_{\mu },  \notag \\
\left[ \widehat{M}_{\mu \nu },\widehat{M}_{\alpha \beta }\right] &=&-\eta
_{\mu \alpha }\widehat{M}_{\nu \beta }+\eta _{\nu \alpha }\widehat{M}_{\mu
\beta }-\eta _{\mu \beta }\widehat{M}_{\alpha \nu }+\eta _{\nu \beta }%
\widehat{M}_{\alpha \mu }.  \label{Fund-PB-2}
\end{eqnarray}%
Hence, $F$ as given by Eq.(\ref{Poincare-generator}) generate respectively
infinitesimal 4-translations (for $b_{\mu \nu }\equiv 0$ and $a_{\mu }\neq 0$%
) and 4-rotations (for $b_{\mu \nu }\neq 0$ and $a_{\mu }\equiv 0$,
corresponding either to Lorentz-boosts or spatial rotations) via
infinitesimal canonical transformations of the type%
\begin{equation}
\mathbf{z}\rightarrow \mathbf{z}^{\prime }+\delta _{o}\mathbf{z},
\end{equation}%
with $\delta _{o}\mathbf{z}\sim O(\delta )$, $\delta >0$ denoting a suitable
infinitesimal. $\delta _{o}\mathbf{z}$ is determined identifying it with$\ $%
\begin{equation}
\delta _{o}\mathbf{z}\equiv \left[ \mathbf{z},F\right] ,
\label{local variation}
\end{equation}%
to be referred to as the \emph{local variation of }$\mathbf{z}$. Hence,
according to DGF\ an arbitrary dynamical variable $\xi $ depending \emph{%
locally }and \emph{smoothly }on the canonical state $\mathbf{z}$ transforms
in terms of the law%
\begin{eqnarray}
&&\left. \xi \rightarrow \xi ^{\prime }=\xi +\delta _{o}\xi ,\right.
\label{variation--law} \\
&&\left. \delta _{o}\xi (\mathbf{z})\equiv \left[ \xi ,F\right] ,\right.
\end{eqnarray}%
where $\delta _{o}\xi (\mathbf{z})=\left[ \xi (\mathbf{z}^{\prime })-\xi (%
\mathbf{z})\right] \left[ 1+O(\delta )\right] .$ It is immediate to
determine an admissible representation for the generators $\left\{ F\right\}
\equiv \left\{ \widehat{p}^{\mu },\widehat{M}^{\mu \nu }\right\} .$ Let us
first consider a relativistic $1$-body system represented by a single
particle in the absence of external forces. For definiteness, let us assume
that $\forall s_{(1)}\in I\equiv
\mathbb{R}
$ the particle 4-velocity is constant. Then $\widehat{p}^{\mu }$ and $%
\widehat{M}^{\mu \nu }$are manifestly given by:
\begin{eqnarray}
\widehat{p}_{\mu } &=&\pi _{\mu }^{\left( 1\right) },  \notag \\
\widehat{M}_{\mu \nu } &=&q_{\mu }^{\left( 1\right) }\pi _{\nu }^{\left(
1\right) }-q_{\nu }^{\left( 1\right) }\pi _{\mu }^{\left( 1\right) },
\end{eqnarray}%
where respectively the 4-vectors $q_{\mu }^{\left( 1\right) }$ and $\pi
_{\mu }^{\left( 1\right) }$ are to be identified with the $1$-body
coordinates and momenta. In the case of the corresponding $N$-body problem
for relativistic interacting particles, in Dirac paper three different
realizations of $\left\{ \widehat{p}^{\mu },\widehat{M}^{\mu \nu }\right\} $
were originally proposed, which are referred to as the instant, point and
front forms. All of them follow by imposing the velocity kinematic
constraints (\ref{VELOCITY KINEMATIC CONSTRAINT}). In particular, the
instant form is realized by prescribing the reference frame in such a way to
set $r_{0}^{\left( i\right) }=0$, for all $i=1,N$, namely describing each
particle position only in terms of the space components $\mathbf{r}^{\left(
i\right) }\equiv r_{l}^{\left( i\right) }$ of its position 4-vector, for $%
l=1,3$. In detail, recalling Eq.(\ref{CANONICAL POSITION}) and introducing
the notation%
\begin{equation}
\pi _{\mu }^{\left( i\right) }=(\pi _{0}^{\left( i\right) },\mathbf{\pi }%
^{\left( i\right) }),
\end{equation}%
according to Dirac the instant form (for $N$-body systems of interacting
particles) is obtained by imposing the velocity kinematic constraints (\ref%
{VELOCITY KINEMATIC CONSTRAINT}) on the \textit{free-particle} canonical
momenta $\pi _{free,\mu }^{\left( i\right) }\equiv m_{o}^{\left( i\right)
}cu_{\mu }^{\left( i\right) }=\left( \pi _{free,0}^{\left( i\right) },%
\mathbf{\pi }_{free}^{\left( i\right) }\right) $ [i.e., in the absence of an
external EM field] such that%
\begin{equation}
\pi _{free,0}^{\left( i\right) }=\sqrt{m_{o}^{\left( i\right) 2}c^{2}+%
\mathbf{\pi }_{free}^{\left( i\right) 2}},  \label{DIRAC-POSITION-1}
\end{equation}%
and then introducing a suitable\emph{\ interaction 4-potential} $V_{\mu
}\equiv (V_{0},\mathbf{V})$ taking into account all the particle
interactions. Letting $l,m=1,3$, this yields the $N$-body \emph{Dirac
constrained instant-form generators }$\left( \widehat{p}_{0},\widehat{p}_{l},%
\widehat{M}_{lm},\widehat{N}_{l0}\right) $ \cite{Dirac}, represented in
terms of the constrained states $\mathbf{z}^{\prime \left( i\right) }=\left(
\mathbf{r}^{(i)},\mathbf{\pi }^{(i)}\right) $ (for $i=1,N$):

\begin{eqnarray}
\widehat{p}_{0} &=&\sum_{i=1,N}p_{0}^{\left( i\right) }=\sum_{i=1,N}\sqrt{%
m_{o}^{\left( i\right) 2}c^{2}+\mathbf{\pi }_{free}^{\left( i\right) 2}}%
+V_{0},  \label{INSTANT-FORM-1} \\
\widehat{p}_{l} &=&\sum_{i=1,N}\pi _{l}^{\left( i\right) }, \\
\widehat{M}_{lm} &=&\sum_{i=1,N}\left[ r_{l}^{\left( i\right) }\pi
_{m}^{\left( i\right) }-r_{m}^{\left( i\right) }\pi _{l}^{\left( i\right) }%
\right] , \\
\widehat{N}_{l0} &=&\sum_{i=1,N}r_{l}^{\left( i\right) }\sqrt{m_{o}^{\left(
i\right) 2}c^{2}+\mathbf{\pi }_{free}^{\left( i\right) 2}}+V_{l},
\label{INSTANT-FORM-4}
\end{eqnarray}%
with $V_{l}\equiv V_{0}\sum_{i=1,N}r_{l}^{\left( i\right) }$ and $V_{0}$
denoting the time-component of a suitable \emph{interaction potential }$4$%
-vector $V_{\mu }\equiv (V_{0},\mathbf{V})$. Here both $\widehat{p}_{0}$ and
$\widehat{N}_{l0}$ are still expressed in terms of the free-particle
canonical momentum $\mathbf{\pi }_{free}^{\left( i\right) }$, while $%
\widehat{N}_{l0}$ differs from $\widehat{M}_{l0}$ because of the imposed
kinematic constraint. Therefore, if interactions occur, their contribution
show up only in $\widehat{p}_{0}$ and $\widehat{N}_{l0}$. In Ref.\cite{Dirac}
the interaction-dependent Poincar\`{e} generators were called
\textquotedblleft Hamiltonians\textquotedblright .

Nevertheless, for the validity of the transformation laws (\ref%
{variation--law}) as well as of the Lorentz conditions (\ref{Fund-PB-2}),
Eqs.(\ref{INSTANT-FORM-1})-(\ref{INSTANT-FORM-4}) are actually to be cast in
terms of the 4-momenta of the interacting system $\mathbf{\pi }^{\left(
i\right) }$ (rather than the free particle momenta $\mathbf{\pi }%
_{free}^{\left( i\right) }$). This means that in general $\mathbf{\pi }%
^{\left( i\right) }$ should be considered as suitably-prescribed functions
of $\mathbf{\pi }_{free}^{\left( i\right) }$ and of the interaction
4-potential $V_{\mu }.$ For definiteness, let us consider\ the case of an
isolated $N$-body system subject only to \emph{binary interactions}
occurring between point particles of the same system. In such a case the
interaction potential 4-vector $V_{\mu }$ is necessarily \emph{separable }%
\cite{Sudarshan1983}, i.e., such that%
\begin{equation}
V_{\mu }\equiv (V_{0},\mathbf{V})=\sum\limits_{i=1,N}V_{\mu }^{(i)},
\label{SEPARABLE 4-potential}
\end{equation}%
with $V_{\mu }^{(i)}$ denoting the $i$\emph{-th particle interaction
potential 4-vector. }Then, assuming that $V_{\mu }^{(i)}$ are only
position-dependent, in view of Eqs.(\ref{INSTANT-FORM-1}) and (\ref%
{INSTANT-FORM-4}), for each particle the canonical 4-momentum of interacting
particles $\pi _{\mu }^{\left( i\right) }$ must depend linearly on $V_{\mu
}^{(i)}$ and $\pi _{free,\mu }^{\left( i\right) },$ namely it takes the form%
\begin{equation}
\pi _{\mu }^{\left( i\right) }=\pi _{free,\mu }^{\left( i\right) }-V_{\mu
}^{(i)},  \label{INTERACTING-P MOMENTA}
\end{equation}%
which implies in turn that necessarily $\pi _{0}^{\left( i\right) }=\sqrt{%
m_{o}^{\left( i\right) 2}c^{2}+\left( \mathbf{\pi }^{\left( i\right) }-%
\mathbf{V}^{(i)}\right) ^{2}}+V_{0}^{(i)}.$ As a consequence, Eqs.(\ref%
{INSTANT-FORM-1})-(\ref{INSTANT-FORM-4}) are actually replaced with
\begin{eqnarray}
\widehat{p}_{0} &=&\sum_{i=1,N}p_{0}^{\left( i\right) }=\sum_{i=1,N}\sqrt{%
m_{o}^{\left( i\right) 2}c^{2}+\left( \mathbf{\pi }^{\left( i\right) }-%
\mathbf{V}^{(i)}\right) ^{2}}+V_{0},  \label{INSTANT-FORM-1-CORRECT} \\
\widehat{p}_{l} &=&\sum_{i=1,N}\pi _{l}^{\left( i\right) },
\label{INSTANT-FORM-2-CORRECT} \\
\widehat{M}_{lm} &=&\sum_{i=1,N}\left[ r_{l}^{\left( i\right) }\pi
_{m}^{\left( i\right) }-r_{m}^{\left( i\right) }\pi _{l}^{\left( i\right) }%
\right] ,  \label{INSTANT-FORM-3-CORRECT} \\
\widehat{N}_{l0} &=&\sum_{i=1,N}r_{l}^{\left( i\right) }\sqrt{m_{o}^{\left(
i\right) 2}c^{2}+\left( \mathbf{\pi }^{\left( i\right) }-\mathbf{V}%
^{(i)}\right) ^{2}}+V_{0i},  \label{INSTANT-FORM-4-CORRECT}
\end{eqnarray}%
where (\ref{INSTANT-FORM-2-CORRECT}) and (\ref{INSTANT-FORM-3-CORRECT})
retain their free-particle form. In order that the Poincar\`{e} generators $%
\widehat{p}_{0}$ and $\widehat{p}_{l}$ commute [in accordance with Eqs.(\ref%
{Fund-PB-2}), with PBs now defined in terms of the constrained state $%
\mathbf{z}^{\prime }$], then it follows necessarily that the 4-vectors $%
V_{\mu }^{(i)},$ for all $i=1,N,$ must be \emph{local} functions of the
4-positions of the particles of the $N$-body system, namely
\begin{equation}
V_{\mu }^{(i)}=V_{\mu }^{(i)}(r_{l}^{\left( 1\right) },...r_{l}^{\left(
N\right) }).  \label{LOCALITY}
\end{equation}%
The explicit proof of this statement is given below [see the subsequent
subsection \#3 and in particular the inequality (\ref{EQUATION FINAL})].
Hence, by construction, in the Dirac approach the \textquotedblleft
Hamiltonians\textquotedblright\ $\widehat{p}_{0}$ and $\widehat{N}_{l0}$ are
\emph{necessarily local functions} too. It follows that DGF\emph{\ applies
only to local Hamiltonian systems. }

It is worth noting that the same conclusion follows directly also from
Dirac's claim (see the quote from his paper given above) that the $N$-body
system should generate a\emph{\ }dynamical system. In fact, in the customary
language of analytical mechanics the latter is intended as a
parameter-dependent map of the phase-space $\Gamma _{N}$ onto itself. This
means that, when the canonical state $\mathbf{z}$ is parametrized in terms
of the coordinate time $t$, there should exist a homeomorphism in $\Gamma
_{N}$ of the form:%
\begin{equation}
\widehat{\mathbf{z}}(t_{o})=\mathbf{z}_{0}\leftrightarrow \widehat{\mathbf{z}%
}(t),  \label{HAMILTONIAN-FLOW}
\end{equation}%
with $t,t_{o}\in I\equiv
\mathbb{R}
.$ Therefore, if the previous statement by Dirac is taken for granted, in
view of the discussion reported above (see Remark \#7 in Section 7), it
implies again that $K_{N}$ must only be a local function of the system
canonical state $\mathbf{z}$.

Let us now analyze, for comparison, the implications of the theory developed
in the present work for EM-interacting $N$-body systems.

\bigskip

\begin{enumerate}
\item \textbf{Non-local Hamiltonian structure}
\end{enumerate}

The first issue is related to the Hamiltonian structure $\left\{ \mathbf{z}%
,K_{N}\right\} $ which characterizes these systems. According to the
Corollary of THM.2 this should be identified with $\left\{ \mathbf{x}%
,H_{N}\right\} .$ Thus, the super-abundant canonical state $\mathbf{z}$
should coincide with $\mathbf{x}\equiv \left\{ \mathbf{x}^{\left( i\right)
},i=1,N\right\} $ spanning the $8N$-dimensional phase-space $\Gamma
_{N}\equiv \Pi _{i=1,N}\Gamma _{1}^{\left( i\right) }$ (with $\Gamma
_{1}^{\left( i\right) }\subset
\mathbb{R}
^{8}$), while $K_{N}$ is identified with the non-local Hamiltonian $%
H_{N}\equiv H_{N}(r,P,\left[ r\right] )$. In particular, $\mathbf{x}^{\left(
i\right) }=\left( r^{\left( i\right) \mu },P_{\mu }^{\left( i\right)
}\right) $ is the corresponding $i$-th particle canonical state, with $%
r^{\left( i\right) \mu }$ and $P_{\mu }^{\left( i\right) }$ denoting
respectively its position and canonical momentum 4-vectors. In the present
case and in contrast to DGF, it follows that:

\begin{enumerate}
\item \emph{Property \#1}: the system Hamiltonian $K_{N}\equiv H_{N}$ must
be necessarily \emph{non-local}.

\item \emph{Property \#2}: the super-abundant state $\mathbf{x}\equiv
\left\{ \mathbf{x}^{\left( i\right) },i=1,N\right\} $ is canonical, namely
it satisfies the canonical equations (\ref{STANDARD HAMILTONIAN FORM}) in
terms of the local PBs defined with respect to the same state. This occurs
if $\mathbf{x}$ is considered unconstrained, i.e., when the $i$-th particle
state $\mathbf{x}^{\left( i\right) }$ is identified with $\mathbf{x}^{\left(
i\right) }=\left( r^{\left( i\right) \mu },P_{\mu }^{\left( i\right)
}\right) $. Hence, the Hamiltonian structure $\left\{ \mathbf{x}%
,H_{N}\right\} $ holds in the unconstrained $8N$-dimensional phase-space $%
\Gamma _{N}\equiv \Pi _{i=1,N}\Gamma _{1}^{\left( i\right) }$, with $\Gamma
_{1}^{\left( i\right) }\subset
\mathbb{R}
^{8}$. As a consequence, also the PBs (including the fundamental PBs (\ref%
{Fund-PB}) and the Lorentz conditions (\ref{Fund-PB-2})) are defined with
respect to the unconstrained state $\mathbf{x}$.

\item \emph{Property \#3}: only the extremal or extremant canonical curves $%
\mathbf{x}(s_{(1)},..s_{(N)})$ [see Eq.(\ref{EXTREMANT CANONICAL-CURVES})]
and not the varied functions satisfy identically the kinematic constraints (%
\ref{VELOCITY KINEMATIC CONSTRAINT}) and (\ref{LINE-ELEMENT
KINEMATIC--CONSTRAINT}).

\item \emph{Property \#4}: the 4-potential $V_{\mu }$ must be necessarily a
non-local function. In particular, for binary EM interactions it must be a
separable function, i.e., of the form (\ref{SEPARABLE 4-potential}):%
\begin{eqnarray}
V_{\mu }(r,\left[ r\right] ) &=&\sum\limits_{i=1,N}V_{\mu }^{(i)}(r,\left[ r%
\right] ), \\
V_{\mu }^{(i)}(r,\left[ r\right] ) &\equiv &\frac{q^{(i)}}{c}A_{\left(
eff\right) \mu }^{(tot)\left( i\right) },
\end{eqnarray}%
with $A_{\left( eff\right) \mu }^{(tot)\left( i\right) }$ being defined by
Eq.(\ref{A-eff-tot}).
\end{enumerate}

\bigskip

\begin{enumerate}
\item[2.] \textbf{Conditions of validity of Dirac instant-form generators}
\end{enumerate}

A further issue is related to the representation of the Poincar\`{e}
generators and, in particular, to the instant form representation given by
Dirac and usually adopted in the literature. The latter is based on\emph{\ }%
Eqs.(\ref{INSTANT-FORM-1})-(\ref{INSTANT-FORM-4}), rather than on Eqs.(\ref%
{INSTANT-FORM-1-CORRECT})-(\ref{INSTANT-FORM-4-CORRECT}), in which\emph{\ }$%
\mathbf{\pi }^{\left( i\right) 2}$ replaces $\mathbf{\pi }_{free}^{\left(
i\right) 2}$ under the square root on the r.h.s. of Eqs.(\ref{INSTANT-FORM-1}%
) and (\ref{INSTANT-FORM-4}). On the other hand, based on the non-local
Hamiltonian structure $\left\{ \mathbf{x},H_{N}\right\} $ expressed in terms
of the unconstrained super-abundant canonical state $\mathbf{x}$ [see Eqs.(%
\ref{CANONICAL STATE}),(\ref{CANONICAL POSITION}) and (\ref{CANONICAL
MOMENTUM})], an admissible realization for $\left\{ \widehat{p}^{\mu },%
\widehat{M}^{\mu \nu }\right\} $ can be determined which holds for an
arbitrary $N\geq 1$. In fact, it is immediate to verify that the
phase-functions%
\begin{eqnarray}
\widehat{p}_{\mu } &=&\sum_{i=1,N}P_{\mu }^{\left( i\right) },  \notag \\
\widehat{M}_{\mu \nu } &=&\sum_{i=1,N}\left[ r_{\mu }^{\left( i\right)
}P_{\nu }^{\left( i\right) }-r_{\nu }^{\left( i\right) }P_{\mu }^{\left(
i\right) }\right] ,  \label{N-BODY-REPRESENTATION}
\end{eqnarray}%
satisfy identically the PBs (\ref{Fund-PB-2}) expressed in terms of the same
state $\mathbf{x}$. In view of Property \#2 this requires that the canonical
generators $\left\{ \widehat{p}^{\mu },\widehat{M}^{\mu \nu }\right\} $
defined by Eq.(\ref{N-BODY-REPRESENTATION}) \emph{must be considered
independent}. Hence, no constraints (on them) can possibly arise by imposing
the validity of the PBs (\ref{Fund-PB-2}).

Another possibility, however, lies in the adoption of a constrained
formulation. This is obtained imposing the kinematic constraints (\ref%
{VELOCITY KINEMATIC CONSTRAINT}) and identifying the canonical state with
the constrained vector $\mathbf{x}^{\prime }\equiv \left( \mathbf{x}^{\prime
(i)},i=1,N\right) $ with $\mathbf{x}^{\prime (i)}=$ $(\mathbf{r}^{(i)},%
\mathbf{P}^{(i)}).$ Recalling again Eqs.(\ref{CANONICAL POSITION}) and (\ref%
{CANONICAL MOMENTUM}), here $\mathbf{r}^{(i)}$ and $\mathbf{P}^{(i)}$ denote
respectively the space parts of the corresponding $i$-th particle 4-vectors.

To carry out a detailed comparison with Dirac, let us consider in particular
the instant-form representation of $\left\{ \widehat{p}^{\mu },\widehat{M}%
^{\mu \nu }\right\} $ as given by Eq.(\ref{N-BODY-REPRESENTATION}). In such
a case the generators are represented by the set, defined for $l,m=1,3$:%
\begin{eqnarray}
\widehat{p}_{0} &=&\sum_{i=1,N}P_{0}^{\left( i\right) },  \label{151} \\
\widehat{p}_{l} &=&\sum_{i=1,N}P_{l}^{\left( i\right) },  \label{152a} \\
\widehat{M}_{lm} &=&\sum_{i=1,N}\left[ r_{l}^{\left( i\right) }P_{m}^{\left(
i\right) }-r_{m}^{\left( i\right) }P_{l}^{\left( i\right) }\right] ,
\label{153a} \\
\widehat{N}_{l0} &=&\sum_{i=1,N}\left[ r_{l}^{\left( i\right) }P_{0}^{\left(
i\right) }\right] ,  \label{155}
\end{eqnarray}%
where $\widehat{p}_{0},$ $\widehat{p}_{l},$ $\widehat{M}_{lm}$ and $\widehat{%
N}_{l0}$ must all be considered as independent. The corresponding \emph{%
constrained }(representation of the)\emph{\ instant-form generators}, with $%
\widehat{p}_{0}$ and $\widehat{N}_{l0}$ expressed in terms of the
constrained state $\mathbf{x}^{\prime },$ become therefore%
\begin{eqnarray}
\left. \widehat{p}_{0}\right\vert _{\mathbf{x}^{\prime }} &=&\sum_{i=1,N}%
\left[ \sqrt{m_{o}^{\left( i\right) 2}c^{2}+\left( \mathbf{P}^{\left(
i\right) }-\frac{q^{\left( i\right) }}{c}\mathbf{A}_{\left( eff\right)
}^{(tot)\left( i\right) }\right) ^{2}}+\frac{q^{\left( i\right) }}{c}%
A_{\left( eff\right) 0}^{(tot)\left( i\right) }\right] ,
\label{Instant-form-1-NEW} \\
\left. \widehat{N}_{l0}\right\vert _{\mathbf{x}^{\prime }} &=&\sum_{i=1,N}%
\left[ r_{l}^{\left( i\right) }\sqrt{m_{o}^{\left( i\right) 2}c^{2}+\left(
\mathbf{P}^{\left( i\right) }-\frac{q^{\left( i\right) }}{c}\mathbf{A}%
_{\left( eff\right) }^{(tot)\left( i\right) }\right) ^{2}}+\frac{q^{\left(
i\right) }}{c}r_{l}^{\left( i\right) }A_{\left( eff\right) 0}^{(tot)\left(
i\right) }\right] ,  \label{Instant-form-2-NEW}
\end{eqnarray}%
where we have represented $A_{\left( eff\right) \mu }^{(tot)\left( i\right)
}\equiv \left( A_{\left( eff\right) 0}^{(tot)\left( i\right) },\mathbf{A}%
_{\left( eff\right) }^{(tot)\left( i\right) }\right) $, with $A_{\left(
eff\right) \mu }^{(tot)\left( i\right) }$ being defined by Eq.(\ref%
{A-eff-tot}) setting $\overline{A}_{\mu }^{(ext)\left( i\right) }=0$. A
characteristic obvious feature of the constrained representations given
above is that of the non-local dependences arising both from binary and self
EM interactions. Analogous conclusions can be drawn also for the so-called
point and front forms of the same generators. This implies that the Lorentz
conditions (\ref{Fund-PB-2}), with PBs now defined in terms of the same
constrained state $\mathbf{x}^{\prime }$, are generally violated. Indeed,
due to the non-locality of $A_{\left( eff\right) \mu }^{(tot)\left( i\right)
}$ in this case the PBs-inequalities%
\begin{equation}
\left[ \widehat{p}_{0\mathbf{x}^{\prime }},\widehat{p}_{l}\right] _{(\mathbf{%
x}^{\prime })}\neq 0  \label{EQUATION FINAL}
\end{equation}%
hold. Hence,\emph{\ if - consistent with DGF - the validity of the Lorentz
conditions (\ref{Fund-PB-2}) is imposed, the constrained forms of the Poincar%
\'{e} generators are manifestly not applicable to the treatment of
EM-interacting }$N$\emph{-body systems.}

Nevertheless, it is immediate to prove that $\left. \widehat{p}%
_{0}\right\vert _{\mathbf{x}^{\prime }}$ indeed generates the correct
evolution equations for the constrained state $\mathbf{x}^{\prime }$. In
fact, denoting by $\left[ \cdot ,\cdot \right] _{(\mathbf{x}^{\prime })}$
the local PBs evaluated with respect to the constrained state\ $\mathbf{x}%
^{\prime }$, let us determine by means of the PBs%
\begin{equation}
\delta _{o}\xi (\mathbf{x})\equiv \left[ \xi ,F\right] _{(\mathbf{x}^{\prime
})},
\end{equation}%
the infinitesimal transformations $\delta _{o}\mathbf{r}^{(i)}$ and $\delta
_{o}\mathbf{P}^{(i)}$ generated by $F=dt\left. \widehat{p}_{0}\right\vert _{%
\mathbf{x}^{\prime }}.$ It is immediate to prove that these yield
respectively%
\begin{eqnarray}
\delta _{o}\mathbf{r}^{(i)\mu } &=&dt\left[ \mathbf{r}^{(i)\mu },\left.
\widehat{p}_{0}\right\vert _{\mathbf{x}^{\prime }}\right] _{(\mathbf{x}%
^{\prime })}\equiv dt\mathbf{v}^{(i)},  \label{DIFFERENCE-0-1} \\
\delta _{o}\mathbf{P}^{(i)} &=&dt\left[ \mathbf{P}^{(i)},\left. \widehat{p}%
_{0}\right\vert _{\mathbf{x}^{\prime }}\right] _{(\mathbf{x}^{\prime
})}\equiv dt\frac{q^{\left( i\right) }}{c}\nabla _{(i)}A_{\left( eff\right)
\nu }^{(tot)\left( i\right) }v^{(i)\nu },  \label{DIFFERENCE-0-2}
\end{eqnarray}%
where the r.h.s. of both equations coincide identically with the spatial
parts of the canonical Eqs.(\ref{DIFFERENCE EQ-1}) and (\ref{DIFFERENCE EQ-2}%
). Hence, as expected, \emph{the constrained state\ }$x^{\prime }$ \emph{is
indeed canonical. Eqs.(\ref{DIFFERENCE-0-1}) and (\ref{DIFFERENCE-0-2})
provide the Hamiltonian equations for }$x^{\prime }$\emph{\ in terms of the
non-local Hamiltonian function} $\left. \widehat{p}_{0}\right\vert _{\mathbf{%
x}^{\prime }}$\emph{. }Again, a necessary and sufficient condition for Eqs.(%
\ref{DIFFERENCE EQ-1}) and (\ref{DIFFERENCE EQ-2}) to hold is that the
fundamental PBs
\begin{eqnarray}
\left[ \mathbf{r}^{\left( i\right) },\mathbf{r}^{\left( j\right) }\right] _{(%
\mathbf{x}_{0}^{\prime })} &=&0,  \notag \\
\left[ \mathbf{P}^{\left( i\right) },\mathbf{P}^{\left( j\right) }\right] _{(%
\mathbf{x}_{0}^{\prime })} &=&0,  \notag \\
\left[ \mathbf{r}^{\left( i\right) },\mathbf{P}^{\left( j\right) }\right] _{(%
\mathbf{x}_{0}^{\prime })} &=&\delta ^{ij}\mathbf{1},  \label{FPB-3}
\end{eqnarray}%
are identically satisfied for all $i,j=1,N$. Here $\mathbf{x}_{0}^{\prime }$
and $\mathbf{x}^{\prime }$ are identified respectively with $\mathbf{x}%
_{o}^{\prime }\equiv \widehat{\mathbf{x}}^{\prime }(t_{o})$ and $\mathbf{x}%
^{\prime }\equiv $ $\widehat{\mathbf{x}}^{\prime }(t)=\mathbf{x}_{o}^{\prime
}+d\mathbf{x}^{\prime }$, with $d\mathbf{x}^{\prime }\mathbf{\equiv }\left(
\delta _{o}\mathbf{x}^{\left( 1\right) },..,\delta _{o}\mathbf{x}^{\left(
N\right) }\right) $, while the previous PBs are evaluated with respect to
the initial state $\mathbf{x}_{o}^{\prime }$. Furthermore, also in this case
the canonical initial history set\emph{\ }$\left\{ \widehat{\mathbf{x}}%
^{\prime }\right\} _{t_{0}}$, to be defined in analogy with Eq.(\ref%
{CANONICAL HISTORY SET}), is assumed prescribed.

\bigskip

\section{\textbf{Non-local generator formalism}}

A basic consequence of the previous considerations is that in the case of
non-local phase-functions, such as $H_{N}$ or $\left. \widehat{p}%
_{0}\right\vert _{\mathbf{x}^{\prime }}$, the local transformation law (\ref%
{variation--law}) becomes inapplicable.

A suitably-modified formulation of DGF appropriate for the treatment of
non-local phase-functions must therefore be developed. This can be
immediately obtained. In fact, let us consider an arbitrary non-local
function of the form $\xi =\xi (\mathbf{z},\left[ \mathbf{z}\right] ),$ with
$\mathbf{z}$ and $\left[ \mathbf{z}\right] $ denoting respectively local and
non-local functional dependences with respect to the canonical state $%
\mathbf{z}$. Let us consider an arbitrary infinitesimal canonical
transformation generated by $F$ of the form $\mathbf{z}\rightarrow \mathbf{z}%
^{\prime }=\mathbf{z}+\left[ \mathbf{z},F\right] $, with $\delta _{o}\mathbf{%
z}$ to be considered as infinitesimal (i.e., of $O(\Delta )$). Then,
requiring that $\xi $ is suitably smooth both with respect to $\mathbf{z}$
and $\left[ \mathbf{z}\right] ,$ the corresponding infinitesimal variation
of $\xi $ can be approximated with
\begin{equation}
\delta \xi (\mathbf{z},\left[ \mathbf{z}\right] )\equiv \left[ \xi (\mathbf{z%
}+\alpha \delta _{o}\mathbf{z},\left[ \mathbf{z}+\alpha \delta _{o}\mathbf{z}%
\right] )-\xi (\mathbf{z},\left[ \mathbf{z}\right] )\right] \left[
1+O(\Delta )\right] ,  \label{APPROXIMATION}
\end{equation}%
$\delta \xi (\mathbf{z},\left[ \mathbf{z}\right] )$ being the (Frechet)
functional derivative of $\xi (\mathbf{z},\left[ \mathbf{z}\right] ),$ namely%
\begin{equation}
\delta \xi (\mathbf{z},\left[ \mathbf{z}\right] )\equiv \lim_{\alpha
\rightarrow 0}\frac{d}{d\alpha }\xi (\mathbf{z}+\alpha \delta _{o}\mathbf{z},%
\left[ \mathbf{z}+\alpha \delta _{o}\mathbf{z}\right] )\equiv \left\{ \xi (%
\mathbf{z},\left[ \mathbf{z}\right] ),F\right\} .
\label{functional-variation}
\end{equation}%
Here $\left\{ \xi (\mathbf{z},\left[ \mathbf{z}\right] ),F\right\} $ denotes
the \emph{non-local Poisson brackets }(NL-PBs) and generally also $F$ can be
considered a non-local function of the form $F\left( \mathbf{z},\left[
\mathbf{z}\right] \right) $ [i.e., of a type analogous to $\xi $]. Such a
definition reduces manifestly to (\ref{variation--law}) in case of local
functions.

Let us prove that the transformation law (\ref{functional-variation}) is
indeed the correct one. To elucidate this point, let us consider the
4-scalar defined by the Dirac-delta $\xi (r,\left[ r\right] )\equiv \delta
\left( \widetilde{R}^{\left( i\right) \alpha }\widetilde{R}_{\alpha
}^{\left( i\right) }-\sigma _{\left( i\right) }^{2}\right) $ entering the
non-local Lagrangian and Hamiltonian functions in the EM self-interaction,
where $\widetilde{R}^{\left( i\right) \alpha }$ denotes the bi-vector
defined by Eq.(\ref{Rtilda}). Let us consider, for example, the action of an
arbitrary infinitesimal Lorentz transformation defined by $\delta
_{o}r^{\left( i\right) \mu }$. In order that $\widetilde{R}^{\left( i\right)
\alpha }\widetilde{R}_{\alpha }^{\left( i\right) }$ is left invariant by the
transformation (Lorentz invariance) it must be%
\begin{equation}
\delta \xi (r,\left[ r\right] )\equiv \left\{ \xi (r,\left[ r\right]
),F\right\} =0,  \label{invariance}
\end{equation}%
with $F=-\widehat{p}^{\mu }a_{\mu }.$ This means that the NL-PBs $\left\{
\xi (r,\left[ r\right] ),F\right\} $ defined by Eq.(\ref%
{functional-variation}), rather than the local PBs $\left[ \xi (r,\left[ r%
\right] ),F\right] ,$ must vanish identically. In particular Eq.(\ref%
{functional-variation}), contrary to the local variation (\ref{local
variation}), preserves the Lorentz invariance of 4-scalars and hence
provides the correct transformations law. Hence, in particular, it follows
that for an isolated $N$-body system with arbitrary $N>1:$
\begin{eqnarray}
\delta H_{N}(r,P,\left[ r\right] ) &\equiv &\left\{ H_{N}(r,P,\left[ r\right]
),F\right\} =0,  \label{LORENTZ-INVARIANCE} \\
\delta \left. \widehat{p}_{0}\right\vert _{\mathbf{x}^{\prime }} &\equiv
&\left\{ \left. \widehat{p}_{0}\right\vert _{\mathbf{x}^{\prime }},F\right\}
=0.  \label{LORENTZ-INVARIANCE-2}
\end{eqnarray}%
It is immediate to prove that Eq.(\ref{LORENTZ-INVARIANCE}) holds by
construction for all Poincar\`{e} generators [see Eqs.(\ref%
{N-BODY-REPRESENTATION}) above], while - instead - generally
\begin{eqnarray}
\delta _{o}H_{N}(r,P,\left[ r\right] ) &\equiv &\left[ H_{N}(r,P,\left[ r%
\right] ),F\right] \neq 0,  \label{VIOLATION OF LORENTZ INV-1} \\
\delta _{o}\left. \widehat{p}_{0}\right\vert _{\mathbf{x}^{\prime }} &\equiv
&\left[ \left. \widehat{p}_{0}\right\vert _{\mathbf{x}^{\prime }},F\right]
_{\left( \mathbf{x}^{\prime }\right) }\neq 0.
\label{VIOLATION OF LORENTZ INV-2}
\end{eqnarray}

Hence, consistent with the results indicated above, we conclude that\emph{\
the local transformation laws realized by the Lorentz conditions (\ref%
{Fund-PB-2}), which are a distinctive feature of DGF, become invalid in the
case of non-local Hamiltonians.}

The \emph{non-local generator formalism} is therefore formally achieved by
imposing modified Lorentz conditions obtained from Eqs.(\ref{Fund-PB-2}), in
which \emph{the local PBs are replaced with the non-local PBs} \emph{defined
by Eq.(\ref{functional-variation})}.

In particular, \emph{the correct transformation laws for the constrained
instant-form Poincar\`{e} generators }[see Eqs.(\ref{152a}),(\ref{153a}) and
(\ref{Instant-form-1-NEW}),(\ref{Instant-form-2-NEW})] \emph{follow by
imposing for the Hamiltonians }[see Eqs.(\ref{Instant-form-1-NEW}) and (\ref%
{Instant-form-2-NEW})]\emph{\ appropriate non-local transformation laws of
the type (\ref{LORENTZ-INVARIANCE-2}), all defined with respect to the
constrained state }$\mathbf{x}^{\prime }$\emph{. }Finally, it must be
remarked that the non-local generator formalism does not affect the validity
of the canonical equations of motion (\ref{DIFFERENCE-0-1}) and (\ref%
{DIFFERENCE-0-2}) as well as the fundamental PBs (\ref{FPB-3}) indicated
above\emph{\ }for the constrained state $\mathbf{x}^{\prime }$, which \emph{%
remain unchanged.}

\bigskip

\section{Counter-examples to the \textquotedblleft
no-interaction\textquotedblright\ theorem}

An open problem in relativistic dynamics is related to the so-called \emph{%
\textquotedblleft no-interaction\textquotedblright\ theorem} due to Currie
(Currie, 1963 \cite{Currie1963}), derived by adopting the DGF, and in
particular the instant form representation for the Poincar\`{e} generators
(see previous Section) given in Ref.\cite{Dirac}. According to this theorem,
an isolated classical $N$-body system of mutually interacting particles
which admits a Hamiltonian structure in which the coordinate variables of
the individual particles coincide with the space parts $3$-vectors of the
particles 4-positions and the canonical equations of motion are Lorentz
covariant, can only be realized by means of a collection of free particles.
This requires, in particular that \emph{\textquotedblleft ...it is
impossible to set up a canonical theory of two interacting particles in
which the individual particle positions are the space parts of
4-vectors\textquotedblright . }In other words, according to the theorem, it
should be impossible to formulate - in terms of a Hamiltonian system - a
covariant canonical theory for an isolated system of $N>1$ classical
particles subject to binary interactions (see also Ref.\cite{Komar1978b}).
The validity of the theorem was confirmed by several other authors (see for
example, Beard and Fong, 1969 \cite{Beard1969}, Kracklauer, 1976 \cite%
{Klacklaure1976}, Martin and Sanz, 1978 \cite{Martin1978}, Mukunda and
Sudarshan, 1981 \cite{Mukunda1981}, Balachandran \textit{et al.}, 1982 \cite%
{Sudarshan1982}). Its original formulation obtained by Currie for the case
of two interacting particles $(N=2)$ was subsequently extended to include
the case $N=3$ (Cannon and Jordan, 1964 \cite{Cannon1964}), first-class
constraints (see Sudarshan and Mukunda, 1983 \cite{Sudarshan1983} and the
corresponding Lagrangian proof given by Marmo \textit{et al.}, 1984 \cite%
{Marmo1984}) and the treatment of curved space-time (De Bi\`{e}vre, 1986
\cite{Bievre1986} and Li, 1989 \cite{Li}). Common assumptions to these
approaches are that:

\begin{enumerate}
\item \emph{Hypothesis \#1:} Both DGF and the Dirac instant form realization
of the Poincar\`{e} generators apply. In particular, the Poincar\`{e}
generators in the instant form, corresponding to the constrained Hamiltonian
structure\ $\left\{ \mathbf{z},K_{N}\right\} ,$ satisfy identically both to
the commutation rules (\ref{Fund-PB-2}) and the kinematic constraints (\ref%
{VELOCITY KINEMATIC CONSTRAINT}).

\item \emph{Hypothesis \#2:} $K_{N}$ admits the Poincar\`{e} group of
symmetry, i.e., it commutes with $\left\{ F\right\} $.

\item \emph{Hypothesis \#3:} All particles, in a suitable proper-time
interval, are not subject to the action of an external force
(locally/globally isolated $N$-body system).
\end{enumerate}

Nevertheless, the theorem has been long questioned (see for example
Fronsdal, 1971 \cite{Fronsdal} and Komar, 1978-1979 \cite%
{Komar1978,Komar1978a,Komar1978b,Komar1979}).\ In particular, there remains
the dilemma whether the \textquotedblleft no-interaction\textquotedblright\
theorem actually applies at all for $N$-body systems subject only to
non-local EM interactions. This refers in particular, to extended charged
particles in the presence of binary and self EM forces. Another interesting
question is whether restrictions placed by the \textquotedblleft
no-interaction\textquotedblright\ theorem actually exist for physically
realizable classical systems. Several authors have advanced the conjecture
that the limitations set by the Currie theorem might be avoided in the
framework of constrained dynamics formulated adopting a
super-abundant-variable canonical approach (see for example Komar, 1978 \cite%
{Komar1978b} and Marmo \textit{et al.}, 1984 \cite{Marmo1984} and references
indicated therein). In particular, to get a better understanding of
interacting $N$-body systems, Todorov \cite{Todorov} and then Komar \cite%
{Komar1978,Komar1978a,Komar1978b,Komar1979} developed a manifestly covariant
classical relativistic model for two particles, of an action-at-a-distance
kind. In the Todorov-Komar model the dynamics is given in terms of two
first-class constraints. An equivalent model was discovered by Droz-Vincent
\cite{Droz,Droz2} based on a two-time formulation of the classical
relativistic dynamics. However, the precise identification of the
Hamiltonian structure $\left\{ \mathbf{z},K_{N}\right\} $ pertaining to $N$%
-body systems subject to EM interactions has remained elusive to date.

Here we claim that counter-examples, escaping both the assumptions and the
restrictions of the \textquotedblleft no-interaction\textquotedblright\
theorem, can be achieved, based on the classical $N$-body system of extended
charged particles formulated here. Starting from the Corollary to THM.2, the
following theorem applies.

\bigskip

\textbf{THM.4 - Standard Hamiltonian form of a locally-isolated }$1$\textbf{%
-body system and a globally-isolated }$N$\textbf{-body system.}

\emph{In validity of THM.2 and of the definitions given by Eqs.(\ref{heff-i}%
)-(\ref{heff-i2}), the following propositions hold:}

\emph{T4}$_{1}$\emph{) The Hamiltonian structure }$\left\{ \mathbf{x}%
,H_{N}\right\} $\emph{\ of the classical system formed by a single extended
charged particle is preserved also in the particular case in which the
external EM }$4-$\emph{potential is such that along the particle world-line}
$r^{(1)}(s_{(1)})$\emph{:}
\begin{equation}
A_{\mu }^{(ext)}(r^{(1)}(s_{(1)}))=\left\{
\begin{array}{ccc}
\neq 0 &  & \forall s_{(1)}\in \left] -\infty ,s_{o}\right] \\
0 &  & \forall s_{(1)}\in \left] s_{o},+\infty ,\right[%
\end{array}%
\right.  \label{world-line conditions}
\end{equation}%
\emph{(locally-isolated particle).}

\emph{T4}$_{2}$\emph{) The Hamiltonian structure }$\left\{ \mathbf{x}%
,H_{N}\right\} $\emph{\ of the classical }$N$\emph{-body system formed by
extended charged particles is preserved also in the particular case in which
the external EM }$4-$\emph{potential vanishes identically,}
\begin{equation}
A_{\mu }^{(ext)}(r)\equiv 0  \label{world-line conditions-2}
\end{equation}%
\emph{(globally-isolated }$N$\emph{-body\ system).}

\emph{Proof} - T4$_{1}$) The proof is an immediate consequence of the
Corollary to THM.2. In fact in the absence of an external EM field, the
effective EM 4-potential $A_{\left( eff\right) \mu }^{(tot)\left( 1\right) }$
(see Eq.(\ref{A-eff-tot})) simply reduces to%
\begin{equation}
A_{\left( eff\right) \mu }^{(tot)\left( 1\right) }=2\overline{A}_{\mu
}^{(self)\left( 1\right) },
\end{equation}%
where, in view of the requirement (\ref{world-line conditions}), $\overline{A%
}_{\mu }^{(self)\left( 1\right) }$ is non-vanishing also in the interval $%
\left] s_{o},+\infty ,\right[ $ (see related discussion in Paper I). Hence,
both the Lagrangian and Hamiltonian equations in standard form [see
respectively Eqs.(\ref{EL-2}) and (\ref{HAM-1}),(\ref{HAM-2})] are
satisfied, with $H_{eff}^{\left( 1\right) }\equiv H_{N,eff}\equiv H_{N}$
still defined by Eqs.(\ref{heff-i2}) and (\ref{SYSTEM HAMILTONIAN}). T4$_{2}$%
) The proof is similar. In this case, due to assumption (\ref{world-line
conditions-2}), $A_{\left( eff\right) \mu }^{(tot)\left( i\right) }$ reduces
to%
\begin{equation}
A_{\left( eff\right) \mu }^{(tot)\left( i\right) }=2\overline{A}_{\mu
}^{(self)\left( i\right) }+\sum_{\substack{ j=1,N  \\ i\neq j}}\overline{A}%
_{\left( eff\right) \mu }^{(bin)\left( ij\right) }.
\end{equation}%
Hence, also in this case both the Lagrangian and Hamiltonian equations in
standard form still hold, with $H_{eff}^{\left( i\right) }$ and $H_{N,eff}$
defined by Eqs.(\ref{heff-i2}) and (\ref{H_N=H_Neff}).

\textbf{Q.E.D.}

\bigskip

It is clear that both propositions T4$_{1}$) and T4$_{2}$) indeed escape the
\textquotedblleft no interaction\textquotedblright\ theorem (avoiding also
the limitations set by its assumptions \#1-\#4). In fact, concerning the
Hamiltonian structure $\left\{ \mathbf{x},H_{N}\right\} $ associated to the
classical $N$-body system of extended charged particles, from THM.4 it
follows that:

\begin{itemize}
\item The effective Hamiltonian is a non-local function of the canonical
state $\mathbf{x}$.

\item The canonical particle equations of motion (\ref{ith-
equation-of-motion}) satisfy the correct transformation laws with respect to
the Poincar\`{e} group, since the non-local system Hamiltonian $H_{N}(r,P,%
\left[ r\right] )$ is by construction a Lorentz 4-scalar.

\item The super-abundant canonical state $\mathbf{x}\equiv \left\{ \mathbf{x}%
^{\left( i\right) },i=1,N\right\} $ is defined in terms of\emph{\ }$\mathbf{x%
}^{\left( i\right) }\equiv \left( r^{\left( i\right) \mu },P_{\mu }^{\left(
i\right) }\right) _{\left( s_{\left( i\right) }\right) },$ where $r^{\left(
i\right) \mu }$ and $P_{\mu }^{\left( i\right) }$ are represented by Eqs.(%
\ref{CANONICAL POSITION}) and (\ref{CANONICAL MOMENTUM}).

\item The extremant curves $\left( \mathbf{x}^{(1)}(s_{(1)}),...\mathbf{x}%
^{(N)}(s_{(N)})\right) $ solutions of Eqs.(\ref{CANONICAL EQUATIONS})
satisfy identically the kinematic constraints (\ref{VELOCITY KINEMATIC
CONSTRAINT}) and (\ref{LINE-ELEMENT KINEMATIC--CONSTRAINT}). As a
consequence, only the space parts of the extremant 4-vectors $r^{\left(
i\right) \mu }(s_{(1)})$ and $P_{\mu }^{\left( i\right) }(s_{(1)})$ are, for
all $i=1,N,$ actually independent.

\item In case of T4$_{1}$), the single-particle motion is non-inertial for
all $s_{(1)}\in I\equiv
\mathbb{R}
.$ Hence, the instant form of Dirac generators [see Eqs.(\ref{INSTANT-FORM-1}%
) and (\ref{INSTANT-FORM-4})] becomes inapplicable even in the case of a $1$%
-body system.
\end{itemize}

\section{On the failure of the \textquotedblleft
no-interaction\textquotedblright\ theorem}

The actual causes of the failure of the \textquotedblleft
no-interaction\textquotedblright\ theorem emerge clearly from the analysis
of the conditions of validity of DGF and the Dirac instant-form generators
(see Section 8). For systems of extended charged particles subject only to
EM interactions the previous assumptions \#1-\#4 (common to all customary
approaches \cite%
{Currie1963,Cannon1964,Beard1969,Klacklaure1976,Martin1978,Mukunda1981,Sudarshan1982,Sudarshan1983,Marmo1984,Bievre1986}%
) which characterize the underlying Hamiltonian structure $\left\{ \mathbf{z}%
,K_{N}\right\} $ make it incompatible with the exact non-local Hamiltonian
structure $\left\{ \mathbf{x},H_{N}\right\} $ determined here. In fact, in
difference to $\left\{ \mathbf{z},K_{N}\right\} $, the Hamiltonian structure
$\left\{ \mathbf{x},H_{N}\right\} $ is characterized by:

\begin{itemize}
\item Super-abundant canonical variables $\mathbf{x}\equiv \left\{ \mathbf{x}%
^{\left( i\right) },i=1,N\right\} ,$ with $\mathbf{x}^{\left( i\right)
}\equiv \left( r^{\left( i\right) \mu },P_{\mu }^{\left( i\right) }\right)
_{\left( s_{\left( i\right) }\right) }$ being the $i$-th particle canonical
state.

\item Extremant curves $\left( \mathbf{x}^{(1)}(s_{(1)}),...\mathbf{x}%
^{(N)}(s_{(N)})\right) $ which satisfy identically the kinematic constraints
discussed above. This is a characteristic property of the canonical
extremant curves only. In fact, the same constraints are not satisfied by
the super-abundant canonical state $\mathbf{x}$.

\item Fundamental PBs (\ref{Fund-PB}) which are satisfied only by the
unconstrained state $\mathbf{x}$. Hence, the non-local Hamiltonian structure
$\left\{ \mathbf{x},H_{N}\right\} $ is warranted if all the canonical
variables defining the state $\mathbf{x}\equiv \left\{ \mathbf{x}^{\left(
i\right) },i=1,N\right\} $ are considered independent. This means that, in
order for the fundamental PBs (\ref{Fund-PB}) to be fulfilled, these
constraints cannot be imposed \textquotedblleft a priori\textquotedblright\
on the canonical state.

\item Poincar\`{e} generators [see Eqs.(\ref{151})-(\ref{155})] which
satisfy the commutation rules (\ref{Fund-PB-2}) when they are considered
independent, as the super-abundant canonical variables $\mathbf{x}\equiv
\left\{ \mathbf{x}^{\left( i\right) },i=1,N\right\} ,$ and fulfilling the
fundamental PBs. For this reason, the Poincar\`{e} generators are
necessarily left unconstrained by imposing the validity of the same
equations [i.e., Eqs.(\ref{Fund-PB-2})].

\item A \emph{non-local} Hamiltonian of the form $H_{N}(r,P,\left[ r\right]
) $. In particular, it follows that $H_{N}$ for classical $N$-body systems
of extended charged particles reduces to a local function only in the case
of \emph{a single isolated particle which exhibits inertial motion.} In view
of THM.1 given in Paper I, this requires the external EM 4-potential acting
on such a particle to vanish identically along the particle world-line,
i.e., $A_{\mu }^{(ext)}(r(s_{1}))=0$ for all $s_{1}\in I\equiv
\mathbb{R}
.$
\end{itemize}

\textit{In conclusion, contrary to the claim of the \textquotedblleft
no-interaction\textquotedblright\ theorem, a Lorentz covariant Hamiltonian
formulation for the dynamics of }$N$\textit{-body systems, with }$N\geq 1$,%
\textit{\ actually exists also for mutually interacting charged particles
subject to binary as well as self EM interactions. The result holds even in
the presence of an external EM field, for extended classical particles
described by the Hamiltonian structure }$\left\{ \mathbf{x},H_{N}\right\} $%
\textit{\ determined here.}

One might conjecture that the validity of the \textquotedblleft
no-interaction\textquotedblright\ theorem could be restored by introducing a
suitable asymptotic approximation for the $N$-body system dynamics. The
latter is related, in particular, to the short delay-time and large-distance
approximations (see Section 7), invoked here for the treatment of particle
self and binary EM interactions. It is immediate to prove that also this
route is necessarily unsuccessful. The reason lays in THM.3 and its
consequences. In fact, as shown above, a Hamiltonian structure of the same
type of $\left\{ \mathbf{x},H_{N}\right\} $ can be recovered for the
asymptotic $N$-body equations of motion determined by the same theorem. This
is identified with the set $\left\{ \mathbf{x},H_{N,eff}^{asym}\right\} $,
where $H_{N,eff}^{asym}=\sum_{i=1,N}H_{eff,asym}^{\left( i\right) }$ and $%
H_{eff,asym}^{\left( i\right) }$ is given by Eq.(\ref{Heff-asym}). By
construction, $\left\{ \mathbf{x},H_{N,eff}^{asym}\right\} $ inherits the
same qualitative properties of the exact Hamiltonian structure $\left\{
\mathbf{x},H_{N}\right\} $. Therefore, in particular, in this approximation $%
\mathbf{x}$ satisfies the fundamental PBs (\ref{Fund-PB}) if it is
unconstrained. In addition, since the same definition applies for the Poincar%
\`{e} generators and their representation in the instant form, the same
conclusions on the validity of the \textquotedblleft
no-interaction\textquotedblright\ theorem follow.

\bigskip

\section{Conclusions}

A formidable open problem in classical mechanics is provided by the missing
consistent Hamiltonian formulation for the dynamics of EM-interacting $N$%
-body systems. This critically affects both classical and quantum mechanics.
In this paper a solution to this fundamental issue has been reached
exclusively within the framework of classical electrodynamics and special
relativity. In particular, the Hamiltonian structure of classical $N$-body
systems composed of EM-interacting finite-size charged particles has been
explicitly determined and investigated.

Both local and non-local EM interactions have been retained. The former are
due to externally-prescribed EM fields, while the latter include both binary
and self EM interactions, both characterized by finite delay-time effects.
Binary interactions occur between any two charges of the $N$-body systems,
while self interactions ascribe to the so-called radiation-reaction
phenomena due to action of the EM self-field\ on a finite-size particle. All
of these contributions have been consistently dealt with in the derivation
of the $N$-body dynamical equations of motion by means of a variational
approach based on the hybrid synchronous Hamilton variational principle.

Both Lagrangian and Hamiltonian covariant differential equations have been
obtained, which are intrinsically of delay-type. The same equations have
also been proved to admit a representation in both standard Lagrangian and
Hamiltonian forms, through the definition of effective non-local Lagrangian
and Hamiltonian functions. The property of Hamilton equations of admitting a
Poisson bracket representation has lead us to prove the existence of a
non-local Hamiltonian structure $\left\{ \mathbf{x},H_{N}\right\} $ for the $%
N$-body system of EM-interacting particles. This has been shown to be
determined by the non-local Hamiltonian function $H_{N}$ and to hold for the
superabundant canonical states $\mathbf{x}$. In particular the correct
Hamiltonian equations of motion are obtained considering the same vector $%
\mathbf{x}$ as unconstrained, the relevant (kinematic) constraints being
satisfied identically by the solution of the same equations.

A further interesting development concerns the asymptotic approximation
determined for the Hamiltonian structure $\left\{ \mathbf{x},H_{N}\right\} $
of the full $N$-body problem. Here we have shown that consistent with the
short delay-time and large-distance asymptotic orderings the latter can be
preserved also by a suitable asymptotic Hamiltonian approximation. In
particular, the perturbative expansion adopted here permits to retain
consistently delay-time contributions, while preserving also the variational
character and the standard Lagrangian and Hamiltonian forms of the $N$-body
dynamical equations. As a basic consequence the very Hamiltonian structure
of the $N$-body problem is warranted. This permits us to overcome the usual
difficulties related to the adoption of non-variational and non-Hamiltonian
approximations previously developed in the literature.

Two important applications of the theory have been pointed out.

The first one concerns the famous and widely cited (both in the context of
classical and quantum mechanics) paper by Dirac (1949) on the generator
formalism approach to the forms of the Poincar\`{e} generators for the
inhomogeneous Lorentz group. Contrary to a widespread belief, we have found
out that the Dirac approach is not valid in the case of $N$-body systems
subject to retarded, i.e., non-local, interactions. In fact, the Lorentz
conditions for the instant-form Poincar\`{e} generators are found to be
satisfied only in the case of local Hamiltonians. Analogous conclusions can
be drawn also for the so-called point and front-forms of the same
generators. Due to the non-local character of the Hamiltonian structure $%
\left\{ \mathbf{x},H_{N}\right\} $ this means that the Dirac generator
formalism expressed in terms of the essential (i.e., constrained) canonical
state $\mathbf{x}^{\prime }$ is not \textquotedblleft \textit{per se}%
\textquotedblright\ directly applicable to the treatment of EM-interacting $%
N $-body systems. However, as shown here, in the same variables its
extension to non-local Hamiltonians can be readily achieved by suitably
modifying the Lorentz conditions so to account for the non-local dependences
of the Hamiltonian structure $\left\{ \mathbf{x},H_{N}\right\} $.

Second, the validity of the Currie \textquotedblleft
no-interaction\textquotedblright\ theorem, concerning the Hamiltonian
description of the relativistic dynamics of isolated interacting particles,
has been investigated. It has been proved that the set $\left\{ \mathbf{x}%
,H_{N}\right\} $ violates the statements of the theorem. The cause of the
failure of theorem (and its proof) lays precisely in the adoption of the
Dirac generator formalism. Explicit counter-examples which overcome the
limitations posed by the \textquotedblleft no-interaction\textquotedblright\
theorem have been issued. Contrary to the claim of the \textquotedblleft
no-interaction\textquotedblright\ theorem, it has been demonstrated that a
standard Hamiltonian formulation for the $N$-body system of charged
particles subject to EM interactions can be consistently formulated.

\acknowledgments The authors are grateful to Marco Meneghini (University of
Trieste, Trieste, Italy) and John C. Miller (SISSA, Trieste, Italy, and
University of Oxford, Oxford, U.K.) for stimulating discussions and helpful
contributions during the initial developments of the paper.

This work was developed in the framework of current PRIN research projects
(2008 and 2009, Italian Ministry of University and Research, Italy) as well
as the research projects of the Consortium for Magnetofluid Dynamics
(University of Trieste, Italy): \textit{Fundamentals and applications of
relativistic Hydrodynamics and Magnetohydrodynamics }(International School
for Advanced Studies (SISSA), Trieste, Italy) and\textit{\
Magnetohydrodynamics in curved space: theory and applications }(Department
of Mathematics and Informatics, University of Trieste, Italy).

\appendix

\section{Evaluation of the action integral of the binary interaction}

In this Appendix the mathematical details of the calculation of the action
integral $S_{C}^{\left( bin\right) \left( i\right) }(r,\left[ r\right] )$
are given. The latter contains the information about the binary EM
interactions among the charged particles of the $N$-body system and it has
been defined by Eqs.(\ref{N-body-funct}) and (\ref{act1ij}) in Section 4-A.
To proceed with the calculation we first notice that, invoking the
definition of the current density given by Eq.(\ref{charge-current-i}), the
functional (\ref{act1ij}) can be equivalently represented as%
\begin{eqnarray}
S_{C}^{(bin)\left( ij\right) }(r,\left[ r\right] ) &=&\frac{q^{\left(
j\right) }}{4\pi \sigma _{\left( j\right) }^{2}c}\int_{1}^{2}d\Omega
A^{(self)\left( i\right) \mu }(r)\int_{-\infty }^{+\infty }ds_{2}\delta
(s_{2}-s_{1\left( j\right) })\times  \notag \\
&&\times \int_{-\infty }^{+\infty }ds_{\left( j\right) }u^{\left( j\right)
\mu }(s_{\left( j\right) })\delta (\left\vert x\left( s_{\left( j\right)
}\right) \right\vert -\sigma _{\left( j\right) })\delta (s_{\left( j\right)
}-s_{2}),  \label{long2ij}
\end{eqnarray}%
where $s_{1\left( j\right) }$ is the root of the equation%
\begin{equation}
u_{\mu }^{\left( j\right) }(s_{1\left( j\right) })\left[ r^{\mu }-r^{\left(
j\right) \mu }\left( s_{1\left( j\right) }\right) \right] =0.
\end{equation}%
Furthermore, because of the principle of relativity, the integral (\ref%
{long2ij}) can be evaluated in an arbitrary reference frame. The explicit
calculation of the integral (\ref{long2ij}) is then achieved, thanks to
Lemma 3 given in Paper I, by invoking a Lorentz boost to the reference frame
$S_{NI}$ moving with 4-velocity $u_{\mu }(s_{2})$. In this frame, by
construction $d\Omega ^{\prime }=cdt^{\prime }dx^{\prime }dy^{\prime
}dz^{\prime }\equiv d\Omega $. In particular, introducing the spherical
spatial coordinates $\left( ct^{\prime },\rho ^{\prime },\vartheta ^{\prime
},\varphi ^{\prime }\right) $ it follows that the transformed spatial volume
element can also be written as $cdt^{\prime }dx^{\prime }dy^{\prime
}dz^{\prime }\equiv cdt^{\prime }d\rho ^{\prime }d\vartheta ^{\prime
}d\varphi ^{\prime }\rho ^{\prime 2}\sin \vartheta ^{\prime }.$ In such a
reference frame the previous scalar equation becomes%
\begin{equation}
u_{\mu }^{\left( j\right) \prime }(s_{1\left( j\right) })\left[ r^{\prime
\mu }-r^{\left( j\right) \prime \mu }\left( s_{1\left( j\right) }\right) %
\right] =0.  \label{bisij}
\end{equation}%
On the other hand, performing the integration with respect to $s_{2}$ in Eq.(%
\ref{long2ij}), it follows that necessarily $s_{2}=s_{1\left( j\right) }$,
so that from Eq.(\ref{bisij}) $s_{1\left( j\right) }$ is actually given by%
\begin{equation}
s_{1\left( j\right) }=ct^{\prime }=s_{2}.
\end{equation}%
As a result, the integral $S_{C}^{(bin)\left( ij\right) }$ reduces to%
\begin{equation}
S_{C}^{(bin)\left( ij\right) }(r^{\prime },\left[ r^{\prime }\right] )=\frac{%
q^{\left( j\right) }}{4\pi \sigma _{\left( j\right) }^{2}c}%
\int_{1}^{2}dx^{\prime }dy^{\prime }dz^{\prime }A^{\left( self\right) \left(
i\right) ^{\prime }\mu }(r^{\prime })\int_{-\infty }^{+\infty }ds_{\left(
j\right) }u^{\left( j\right) \prime \mu }(s_{\left( j\right) })\delta
(\left\vert x^{\left( j\right) \prime }\left( s_{\left( j\right) }\right)
\right\vert -\sigma _{\left( j\right) }),
\end{equation}%
with $x^{\left( j\right) \prime \mu }\left( s_{\left( j\right) }\right)
=r^{\prime \mu }-r^{\left( j\right) \prime \mu }\left( s_{\left( j\right)
}\right) $. Moreover%
\begin{equation}
A_{\mu }^{\left( self\right) \left( i\right) \prime }(r^{\prime
})=2q^{\left( i\right) }\int_{-\infty }^{+\infty }ds_{\left( i\right)
}^{\prime \prime }u_{\mu }^{\left( i\right) \prime }\left( s_{\left(
i\right) }^{\prime \prime }\right) \delta (\widehat{R}^{\left( i\right)
\prime \alpha }\widehat{R}_{\alpha }^{\left( i\right) \prime }),
\end{equation}%
with $\widehat{R}^{\left( i\right) \prime \alpha }=r^{\prime \alpha
}-r^{\left( i\right) \prime \alpha }(s_{\left( i\right) }^{\prime \prime })$.

Hence, $S_{C}^{(bin)\left( ij\right) }$ reduces to the functional form:%
\begin{eqnarray}
S_{C}^{(bin)\left( ij\right) }(r^{\prime },\left[ r^{\prime }\right] ) &=&%
\frac{2q^{\left( i\right) }q^{\left( j\right) }}{4\pi \sigma _{\left(
j\right) }^{2}c}\int_{0}^{\pi }d\vartheta ^{\prime }\sin \vartheta ^{\prime
}\int_{0}^{2\pi }d\varphi ^{\prime }\int_{0}^{+\infty }d\rho ^{\prime }\rho
^{^{\prime }2}\times  \notag \\
&&\times \int_{-\infty }^{+\infty }ds_{\left( i\right) }^{\prime \prime
}u_{\mu }^{\left( i\right) \prime }\left( s_{\left( i\right) }^{\prime
\prime }\right) \delta (\widehat{R}^{\left( i\right) \prime \alpha }\widehat{%
R}_{\alpha }^{\left( i\right) \prime })\int_{-\infty }^{+\infty }ds_{\left(
j\right) }u^{\left( j\right) \prime \mu }(s_{\left( j\right) })\delta
(\left\vert x^{\left( j\right) \prime }\left( s_{\left( j\right) }\right)
\right\vert -\sigma _{\left( j\right) }).
\end{eqnarray}%
The remaining spatial integration can now be performed letting%
\begin{equation}
\rho ^{\prime }\equiv \left\vert x^{\prime }\left( s_{\left( j\right)
}\right) \right\vert
\end{equation}%
and making use of the spherical symmetry of the charge distribution. The
constraints placed by the two Dirac-delta functions $\delta (\widehat{R}%
^{\left( i\right) \prime \alpha }\widehat{R}_{\alpha }^{\left( i\right)
\prime })$ and $\delta (\left\vert x^{\prime }\left( s_{\left( j\right)
}\right) \right\vert -\sigma _{\left( j\right) })$ in the previous equation
imply that both $\widehat{R}^{\left( i\right) \prime \alpha }\widehat{R}%
_{\alpha }^{\left( i\right) \prime }$ and $\left\vert x^{\prime }\left(
s_{\left( j\right) }\right) \right\vert $ are 4-scalars. Then, introducing
the representation%
\begin{equation}
\widehat{R}^{\left( i\right) \prime \alpha }\equiv r^{\prime \alpha
}-r^{\left( i\right) \prime \alpha }(s_{\left( i\right) }^{\prime \prime })=%
\widetilde{R}^{\left( ij\right) \prime \alpha }+x^{\left( j\right) \prime
\alpha }\left( s_{j}\right) ,
\end{equation}%
with%
\begin{eqnarray}
\widetilde{R}^{\left( ij\right) \prime \alpha } &\equiv &r^{\left( j\right)
\prime \alpha }\left( s_{\left( j\right) }\right) -r^{\left( i\right) \prime
\alpha }(s_{\left( i\right) }^{\prime \prime }), \\
x^{\left( j\right) \prime \alpha }\left( s_{j}\right) &\equiv &r^{\prime
\alpha }-r^{\left( j\right) \prime \alpha }\left( s_{\left( j\right)
}\right) ,
\end{eqnarray}%
it follows that%
\begin{equation}
\widehat{R}^{\left( i\right) \prime \alpha }\widehat{R}_{\alpha }^{\left(
i\right) \prime }=\widetilde{R}^{\left( ij\right) \prime \alpha }\widetilde{R%
}_{\alpha }^{\left( ij\right) \prime }+x^{\left( j\right) \prime \alpha
}\left( s_{\left( j\right) }\right) x_{\alpha }^{\left( j\right) \prime
}\left( s_{\left( j\right) }\right) +2\widetilde{R}^{\left( ij\right) \prime
\alpha }x_{\alpha }^{\left( j\right) \prime }\left( s_{\left( j\right)
}\right)
\end{equation}%
is necessarily a 4-scalar independent of the integration angles $\left(
\varphi ^{\prime },\vartheta ^{\prime }\right) $ when evaluated on the
hypersurface $\Sigma :\widehat{R}^{\left( i\right) \prime \alpha }\widehat{R}%
_{\alpha }^{\left( i\right) \prime }=0$. Similarly, the Dirac-delta $\delta
(\left\vert x^{\left( j\right) \prime }\left( s_{\left( j\right) }\right)
\right\vert -\sigma _{\left( j\right) })$ warrants that $x^{\left( j\right)
\prime \alpha }\left( s_{\left( j\right) }\right) x_{\alpha }^{\left(
j\right) \prime }\left( s_{\left( j\right) }\right) =-\sigma _{\left(
j\right) }^{2},$ which is manifestly a 4-scalar too. Let us now prove that
necessarily
\begin{equation}
\widetilde{R}^{\left( ij\right) \prime \alpha }x_{\alpha }^{\left( j\right)
\prime }\left( s_{\left( j\right) }\right) \equiv 0.  \label{sym0-ij}
\end{equation}%
In fact, on $\Sigma $ it must be%
\begin{eqnarray}
\frac{d}{ds_{\left( j\right) }}\left[ \widehat{R}^{\left( i\right) \prime
\alpha }\widehat{R}_{\alpha }^{\left( i\right) \prime }\right] &=&\frac{d}{%
ds_{\left( i\right) }^{\prime \prime }}\left[ \widehat{R}^{\left( i\right)
\prime \alpha }\widehat{R}_{\alpha }^{\left( i\right) \prime }\right] =0,
\label{sym1-ij} \\
\frac{d}{ds_{\left( j\right) }}\left[ \widetilde{R}^{\left( ij\right) \prime
\alpha }x_{\alpha }^{\left( j\right) \prime }\left( s_{\left( j\right)
}\right) \right] &=&u^{\left( j\right) \prime \alpha }\left( s_{\left(
j\right) }\right) x_{\alpha }^{\left( j\right) \prime }\left( s_{\left(
j\right) }\right) -\widetilde{R}^{\left( ij\right) \prime \alpha }u_{\alpha
}^{\left( j\right) \prime }\left( s_{\left( j\right) }\right) =  \notag \\
&=&-\widetilde{R}^{\left( ij\right) \prime \alpha }u_{\alpha }^{\left(
j\right) \prime }\left( s_{\left( j\right) }\right) =-\frac{1}{2}\frac{d}{%
ds_{\left( j\right) }}\left[ \widetilde{R}^{\left( ij\right) \prime \alpha }%
\widetilde{R}_{\alpha }^{\left( ij\right) \prime }\right] , \\
\frac{d}{ds_{\left( i\right) }^{\prime \prime }}\left[ \widetilde{R}^{\left(
ij\right) \prime \alpha }x_{\alpha }^{\left( j\right) \prime }\left(
s_{\left( j\right) }\right) \right] &=&-u^{\left( i\right) \prime \alpha
}\left( s_{\left( i\right) }^{\prime \prime }\right) x_{\alpha }^{\left(
j\right) \prime }\left( s_{\left( j\right) }\right) , \\
\frac{d}{ds_{\left( i\right) }^{\prime \prime }}\left[ \widetilde{R}^{\left(
ij\right) \prime \alpha }\widetilde{R}_{\alpha }^{\left( ij\right) \prime }%
\right] &=&-2\widetilde{R}^{\left( ij\right) \prime \alpha }u_{\alpha
}^{\left( i\right) \prime }\left( s_{\left( i\right) }^{\prime \prime
}\right) .
\end{eqnarray}%
Therefore,%
\begin{equation}
\frac{d}{ds_{\left( j\right) }}\left[ \widehat{R}^{\left( i\right) \prime
\alpha }\widehat{R}_{\alpha }^{\left( i\right) \prime }\right] =\frac{d}{%
ds_{\left( j\right) }}\left[ \widetilde{R}^{\left( ij\right) \prime \alpha }%
\widetilde{R}_{\alpha }^{\left( ij\right) \prime }+2\widetilde{R}^{\left(
ij\right) \prime \alpha }x_{\alpha }^{\left( j\right) \prime }\left(
s_{\left( j\right) }\right) \right] =0,
\end{equation}%
\begin{eqnarray}
\frac{d}{ds_{\left( i\right) }^{\prime \prime }}\left[ \widehat{R}^{\left(
i\right) \prime \alpha }\widehat{R}_{\alpha }^{\left( i\right) \prime }%
\right] &=&\frac{d}{ds_{\left( i\right) }^{\prime \prime }}\left[ \widetilde{%
R}^{\left( ij\right) \prime \alpha }\widetilde{R}_{\alpha }^{\left(
ij\right) \prime }+2\widetilde{R}^{\left( ij\right) \prime \alpha }x_{\alpha
}^{\left( j\right) \prime }\left( s_{\left( j\right) }\right) \right] =
\notag \\
&=&-2\widetilde{R}^{\left( ij\right) \prime \alpha }u_{\alpha }^{\left(
i\right) \prime }\left( s_{\left( i\right) }^{\prime \prime }\right)
-2u^{\left( i\right) \prime \alpha }\left( s_{\left( i\right) }^{\prime
\prime }\right) x_{\alpha }^{\left( j\right) \prime }\left( s_{\left(
j\right) }\right) =0,
\end{eqnarray}%
from which it follows that, on $\Sigma $, $\widetilde{R}^{\left( ij\right)
\prime \alpha }$ is a 4-vector, since by definition both $u_{\alpha
}^{\left( i\right) \prime }\left( s_{\left( i\right) }^{\prime \prime
}\right) $ and $x_{\alpha }^{\left( j\right) \prime }\left( s_{\left(
j\right) }\right) $ are 4-vectors too. Now we notice that%
\begin{equation}
\widetilde{R}^{\left( ij\right) \prime \alpha }\widetilde{R}_{\alpha
}^{\left( ij\right) \prime }=f\left( s_{\left( j\right) },s_{\left( i\right)
}^{\prime \prime }\right) =f\left( s_{\left( i\right) }^{\prime \prime
},s_{\left( j\right) }\right) ,  \label{sym2-ij}
\end{equation}%
with $f$ being a 4-scalar which is symmetric with respect to $s_{\left(
j\right) }$ and $s_{\left( i\right) }^{\prime \prime }$, while by
construction%
\begin{equation}
\widetilde{R}^{\left( ij\right) \prime \alpha }x_{\alpha }^{\left( j\right)
\prime }\left( s_{\left( j\right) }\right) =g\left( s_{\left( j\right)
},s_{\left( i\right) }^{\prime \prime },\sigma _{\left( j\right) }\right)
\neq g\left( s_{\left( i\right) }^{\prime \prime },s_{\left( j\right)
},\sigma _{\left( i\right) }\right) ,  \label{sym3-ij}
\end{equation}%
where $g$ is a non-symmetric 4-scalar with respect to the same parameters.
On the other hand, Eq.(\ref{sym1-ij}) requires that $\widehat{R}^{\left(
i\right) \prime \alpha }\widehat{R}_{\alpha }^{\left( i\right) \prime }$
must be symmetric in both $s_{\left( j\right) }$ and $s_{\left( i\right)
}^{\prime \prime }$, so that, thanks to Eqs.(\ref{sym2-ij}) and (\ref%
{sym3-ij}), we can conclude that $g$ is a constant 4-scalar. To determine
the precise value of $g=\widetilde{R}^{\left( ij\right) \prime \alpha
}x_{\alpha }^{\left( j\right) \prime }\left( s_{\left( j\right) }\right) $
we evaluate it in the $j$-th particle COS co-moving reference frame, where
by definition $r_{COS}^{\left( j\right) \mu }\left( s_{0}\right) =\left(
s_{0},\mathbf{0}\right) $ for all the COS proper times $s_{0}\in \lbrack
-\infty ,+\infty ]$. In this frame$\widetilde{R}^{\left( ij\right) \prime
\alpha }=\left( s_{\left( j\right) }-s_{\left( i\right) }^{\prime \prime },%
\mathbf{0}\right) $ has only time component and when $s_{0}=s_{\left(
j\right) }$ we get $g=\widetilde{R}^{\left( ij\right) \prime \alpha
}x_{\alpha }^{\left( j\right) \prime }\left( s_{\left( j\right) }\right) =0$
identically. On the other hand, since $g$ is a 4-scalar, it is independent
of both $s_{\left( j\right) }$ and $s_{\left( i\right) }^{\prime \prime }$
and it is null when $s_{0}=s_{\left( j\right) }$, we conclude that it must
be null for all $s_{0}$ and in any reference frame, which proves Eq.(\ref%
{sym0-ij}).

Hence, as a result of the integration, the action integral $%
S_{C}^{(bin)\left( ij\right) }$ carrying the interaction of particle $i$ on
particle $j$ takes necessarily the expression%
\begin{equation}
S_{C}^{(bin)\left( ij\right) }(r^{\prime },\left[ r^{\prime }\right] )=\frac{%
2q^{\left( i\right) }q^{\left( j\right) }}{c}\int_{1}^{2}dr_{\mu }^{\left(
i\right) \prime }\left( s_{\left( i\right) }^{\prime \prime }\right)
\int_{1}^{2}dr^{\left( j\right) \prime \mu }(s_{\left( j\right) }^{\prime
})\delta (\widetilde{R}^{\left( ij\right) \prime \alpha }\widetilde{R}%
_{\alpha }^{\left( ij\right) \prime }-\sigma _{\left( j\right) }^{2}).
\end{equation}%
Finally, since by construction $S_{C}^{(bin)\left( ij\right) }$ is a
4-scalar, by dropping the primes and replacing $s_{\left( i\right) }^{\prime
\prime }$ and $s_{\left( j\right) }^{\prime }$ respectively with $s_{\left(
i\right) }$ and $s_{\left( j\right) }$, the result reported in Eq.(\ref%
{act1ij-bis}) is recovered.

\end{document}